\documentclass{aa}
\usepackage{amsmath}
\usepackage{placeins}
\usepackage{wrapfig}
\usepackage{array}
\usepackage{afterpage}
\usepackage[usenames,dvipsnames]{xcolor}
\usepackage{multirow}
\usepackage{float}

\newcommand{\chemcomp}{\texttt{chemcomp} }
\newcommand{\low}[1]{_{\text{#1}}}
\newcommand{\high}[1]{^{\text{#1}}}

\makeatletter
\renewcommand*\aa@pageof{, page \thepage{} of \pageref*{LastPage}}
\makeatother

\begin{document}
\setlength\extrarowheight{1.5pt}

\title{Disk and atmosphere composition of multi-planet systems}

\author{Mark Eberlein\inst{1} \and Bertram Bitsch\inst{2,3}
\and Ravit Helled\inst{1}}

\institute{
    Department of Astrophysics, University of Zurich, Winterthurerstrasse 190, 8057 Zurich, Switzerland
    \and
    Department of Physics, University College Cork, Cork, Ireland
    \and
    Max-Planck-Institut für Astronomie, Königstuhl 17, 69117 Heidelberg, Germany}

\abstract{
In protoplanetary disks, small mm-cm-sized pebbles drift inwards which can aid planetary growth and influence the chemical composition of their natal disks. 
Gaps in protoplanetary disks can hinder the effective inward transport of pebbles by trapping the material in pressure bumps.
Here we explore how multiple planets change the vapour enrichment by gap opening. For this, we extend the \chemcomp code to include multiple growing planets and investigate the effect of 1, 2 \& 3 planets on the water content and C/O ratio in the gas disk as well as the final composition of the planetary atmosphere. 
We follow planet migration over evaporation fronts and find that previously trapped pebbles evaporate relatively quickly and enrich the gas. 
We also find that in a multi-planet system, the atmosphere composition can be reduced in carbon and oxygen compared to the case without other planets, due to the blocking of volatile-rich pebbles by an outer planet. This effect is stronger for lower viscosities because planets migrate further at higher viscosities and eventually cross inner evaporation fronts, releasing the previously trapped pebbles. Interestingly, we find that nitrogen remains super-stellar regardless of the number of planets in the system such that super-stellar values in N/H of giant planet atmospheres may be a tracer for the importance of pebble drift and evaporation.
}


\maketitle

\section{Introduction} \label{sec:intro}
Planets form in disks where multi-planetary systems are formed. Indeed, many of the observed planetary systems consist of more than one planet \citep{mayor_harps_2011, rosenthal_california_2022} and many of the observed disks show substructures \citep{ansdell_alma_2016, andrews_disk_2018, garufi_sphere_2024} that might be explained by planets \citep[e.g.][]{bae_structured_2023} as was confirmed in the case of the disk PDS 70 \citep{keppler_discovery_2018, muller_orbital_2018, benisty_circumplanetary_2021}.
Therefore planet formation theory should account for the formation of several planets \citep{bitsch_formation_2019, matsumura_n-body_2021, izidoro_formation_2021, emsenhuber_new_2021} and explore how their interaction affects the final planetary (and disk) properties. 
In this study, we consider the formation of a few giant planets including gap formation and investigate how this affects the enrichment of the disk and planets focusing on pebble flux.

The mechanism for giant planet formation is still being investigated with the two leading models being core accretion \citep{pollack_formation_1996, ormel_effect_2010, lambrechts_rapid_2012} and disk instability \citep{boss_giant_1997, boss_evolution_1998}.  While the composition of giant planets is dominated by hydrogen and helium, they are also expected to include heavy elements which can be accreted as solids or gas.    
Solid material can either be accreted in the form of small mm-cm-sized pebbles or in the form of larger planetesimals. Both can change the heavy element enrichment of a growing planet \citep[e.g.][]{knierim_constraining_2022, schneider_how_2021-1}. Pebbles experience a strong gas drag resulting in inward movement into hotter regions and evaporate \citep{weidenschilling_aerodynamics_1977, brauer_coagulation_2008, birnstiel_gas-_2010, booth_planet-forming_2019, schneider_how_2021-1, molliere_interpreting_2022, mah_close-ice_2023}. This changes the chemical composition of the disk depending on the distance to the star and therefore the composition of the material that a forming planet can accrete. However, planets that open gaps can significantly reduce the pebble flux by blocking them outside their pressure bump. This alters the gas composition of the disk harbouring a planet by reducing the amount of pebbles evaporating at different ice lines \citep{schneider_how_2021-1, kalyaan_effect_2023}.
Previous studies mostly focused on how a single planet influences the composition of a disk \citep[e.g.][]{schneider_how_2021-1} or how pebble evaporation influences the composition of planets \citep[e.g.][]{schneider_how_2021-1, schneider_how_2021, bitsch_how_2022, kalyaan_effect_2023}, but so far no study included the effects of multiple planets and investigated the influence on the atmospheric composition.

Here we investigate how multiple planets influence the water content of the inner disk and the bulk atmospheric composition depending if one, two or three planets form in the same disk. We use the \chemcomp code that models planet formation in a viscous accretion disk first proposed in \citet{schneider_how_2021-1} and present the changes required to extend the code to multiple planets.

The paper is structured as follows, in sec. \ref{sec:method} we explain briefly the key ideas of the model and explain in detail the changes to the previous code. In sec. \ref{sec:results} we show the results of our simulations and in sec. \ref{sec:ModelLimitations} discuss the model limitations. We summarise and conclude the main findings in sec. \ref{sec:summary}.
The appendix discusses the pebble size limits (sec. \ref{sec:PebbleSizeLimits}), water vapour transport (sec. \ref{sec:WaterTransport}), disk nitrogen abundance (sec. \ref{sec:DiskNitrogenAbundance}) and contains tables on the final planet composition and mass (sec. \ref{sec:FinalPlanetCompositions}), condensation temperature and volume mixing ratios of molecules (sec. \ref{sec:CondensationTAndVolMixRatios}) and the disk composition (sec. \ref{sec:InitialDiskComposition}).

\section{Method} \label{sec:method}
To model the viscous evolution of a protoplanetary disk and the formation of a gas giant the 1D semi-analytic code \chemcomp \citep{schneider_how_2021-1} (made publicly available in \citet{schneider_chemcomp_2023}), is adapted to host multiple planets in the same disk. For simplicity, we don't include the direct gravitational force between the planets but changes in the disk composition affect the accreted material of other planets in the same system. We explore how the growth and gap formation of multiple planets influence their chemical composition and that of their birth disk.

\subsection{Disk evolution}
The disk evolution is based on the $\alpha$-prescription \citep{shakura_black_1973} to account for small-scale turbulence by varying the viscosity $\nu$. Given the isothermal sound speed $c\low{s}$ and the Kepler orbital period $\Omega\low{K}=\sqrt{GM/r^3}$ at distance $r$ from the star the viscosity is given by:
\begin{equation}
    \nu=\alpha\frac{c^2\low{s}}{\Omega\low{K}}
\end{equation}
We vary the numerical parameter $\alpha$ within the range of $10^{-4} \leq \alpha \leq 10^{-3}$. Tab. \ref{tab:default_setup} lists all the relevant input parameters to our model and their values. The viscous gas evolution equation can be derived by considering the conservation of mass and angular momentum throughout the disk \citep{lynden-bell_evolution_1974}.
We use the equation for every molecular species Y, which can be written as
\begin{equation}\label{eq:viscous-evolution}
    \frac{\partial\Sigma_\text{gas,Y}}{\partial t} - \frac{3}{r}\frac{\partial}{\partial r}\left[\sqrt{r}\frac{\partial}{\partial r}\left(\sqrt{r}\nu\Sigma_\text{gas,Y}\right)\right]=\dot{\Sigma}_\text{Y},
\end{equation}
where $\Sigma_\text{gas, Y}$ is the gas surface density and $\dot{\Sigma}_\text{Y}$ is a source term. 

The equations are solved on a logarithmic grid with time steps of $\Delta t=10$ yr based on the algorithm for the advection-diffusion problem in \citet{birnstiel_gas-_2010}. All quantities are defined on a grid with $N\low{grid}=5000$ grid points spanning over a range from $r\low{in}=0.1$  AU to $r\low{out}=1000$ AU with an inner Neumann boundary condition for the gas and solid surface densities and an outer Dirichlet condition.
This grid resolution is sufficiently high to follow the disk's evolution properly.
We also ensured that the disk's physical properties don't change depending on a planet being close to a cell boundary. 
The initial profile is set to an analytic solution \citep{lin_tidal_1986, lodato_protoplanetary_2017}.

The disk's mid-plane temperature is calculated using an iterative approach by considering the stellar flux, the heat generated by viscosity and the vertical heat diffusion throughout the disk. The initial guess for the temperature profile only considers the irradiation of the central star and is therefore a simple power law which depends on the distance to the star. For details see \citet[Appendix B]{schneider_how_2021-1}.
We keep the temperature profile fixed over time. The disk dispersal at the end of its lifetime is modelled by reducing the gas surface density exponentially on a time scale $\tau\low{decay}$ after $t\low{evap}$ years \citep[see eq. (20)][]{schneider_how_2021-1}.

For dust and pebbles, we use the two-population \citep{birnstiel_simple_2012} approach, where dust particles are represented by one size and pebbles by another. We considered the growth of pebbles limited by three factors, namely the drift limit where inward drift due to gas drag prevents bigger particles, the fragmentation limit, where pebbles fragment due to collisions and the drift-induced fragmentation depending on the relative drift speed of the particles (see sec. \ref{sec:PebbleSizeLimits} for more details). The fragmentation velocity is set to $v\low{f}=5$ m/s for the entire disk, based on the average value of laboratory experiments that give values between $1-10$ m/s depending on the pebble's composition \citep[e.g.][]{gundlach_stickiness_2015}.

Depending on the evaporation temperature of a molecular species, the material contained in pebbles can evaporate and be added to the gas or vice versa in the case of condensation. The evaporation timescale is taken to be such that every particle that crosses an evaporation line evaporates exponentially over a region of $1\times10^{-3}$ AU.
Note that the exact position of evaporation is limited by the cell size. We therefore considered different grid resolutions and found that the disk evolution is similar. 
We calculate the condensation rate by considering the mass that can stick to already existing pebbles or dust particles. I.e. material crossing an ice line from either inward or outward is not evaporated or condensed instantaneously. To ensure mass conservation within a finite time step the condensation and evaporation per time step is limited to 90\% of the local surface density.

We assume a solar composition for the disk as given by \citet{asplund_chemical_2009} (see Tab. \ref{tab:solar_elemental_abundances}) (in the following, we refer to the composition as stellar rather than solar composition).
The volume mixing ratios, defining the ratios of different molecules, are based on the 20\% carbon grain model of \citet{schneider_how_2021-1} with evaporation temperatures from \citet{lodders_solar_2003}. Tab. (\ref{tab:mixing_ratios}) in the appendix shows the volume mixing ratios and the evaporation temperatures.

\subsection{Planet formation}
For our model, we divide planet formation into two phases, following the core accretion model \citep{pollack_formation_1996}. The first phase starts with a planetary embryo with the pebble transition mass \citep{lambrechts_rapid_2012}, where pebble accretion becomes efficient. Reaching the pebble isolation mass \citep{lambrechts_separating_2014,bitsch_pebble-isolation_2018} marks the start of the second phase. 

The pebble accretion rate is calculated depending on the accretion regime either in a 2D or 3D manner \citep{morbidelli_great_2015, johansen_forming_2017} and accreted pebbles are removed from the disk. We model a simple primordial atmosphere formation during this phase by inserting 10\% of the accreted solids inside the atmosphere instead of the core following our previous approaches. Note we do not model any atmosphere structure.

If the planet reaches the pebble isolation mass, pebble accretion stops and the planet starts accreting gas. Because different processes can govern the accretion of gas, we use the minimum of three accretion rates, namely the Ikoma accretion rate that happens on the Kelvin-Helmholtz timescale \citep{ikoma_formation_2000}, the Machida accretion rate \citep{machida_gas_2010} based on 3D hydrodynamical shearing box simulations and the accretion rate provided by the mass flux through the disk. The Machida accretion rate is scaled by the surface density averaged over the Hill region, which we define for our purpose as the annulus spanning from $r=a\low{p}-r\low{H}$ to $r=a\low{p}+r\low{H}$, with $a\low{p}$ and $r\low{H}$ being the planet's semi-major axis and Hill radius respectively. Contrary to the original code the accreted gas onto the planet is now removed from the disk. The accretion is spread over a continuous profile $f\low{red}$ \citep{bergez-casalou_influence_2020} that depends on the distance $d$ to the planet relative to its Hill radius. With $x=d/r\low{H}$ the profile reads
\begin{equation}
    f\low{red}\!=\!
    \begin{cases}
        2/3& x<0.45\\
        2/3\times\cos^4\left(\pi\left(x-0.45\right)\right)&0.45<x<0.9
    \end{cases}
\end{equation}
The gas surface density $\Sigma\low{acc,Y}$ of a given molecular species Y is then reduced in a cell with a cell centre at a distance $r_i$ to the star by using the following equation
\begin{equation}
    \dot{\Sigma}\low{gas,Y}\high{}\left(r_i\right)=-f\low{red}\left(\frac{r_i-a\low{p}}{r\low{H}}\right)\times\frac{\dot{M}\low{gas,p}}{2\pi r_i\Delta r_i}
\end{equation}
where $a\low{p}$ is the planet's semi-major axis, $\dot{M}\low{gas,p}$ the gas accretion rate of the planet and $\Delta r_i$ the size of the respective cell at position $r_i$. To ensure mass conservation on a finite grid the final removal of the mass is again normalised to match $\dot{M}\low{gas,p}$.

\subsection{Gap opening}
\begin{figure}[ht!]
    \includegraphics[width=0.98\linewidth]{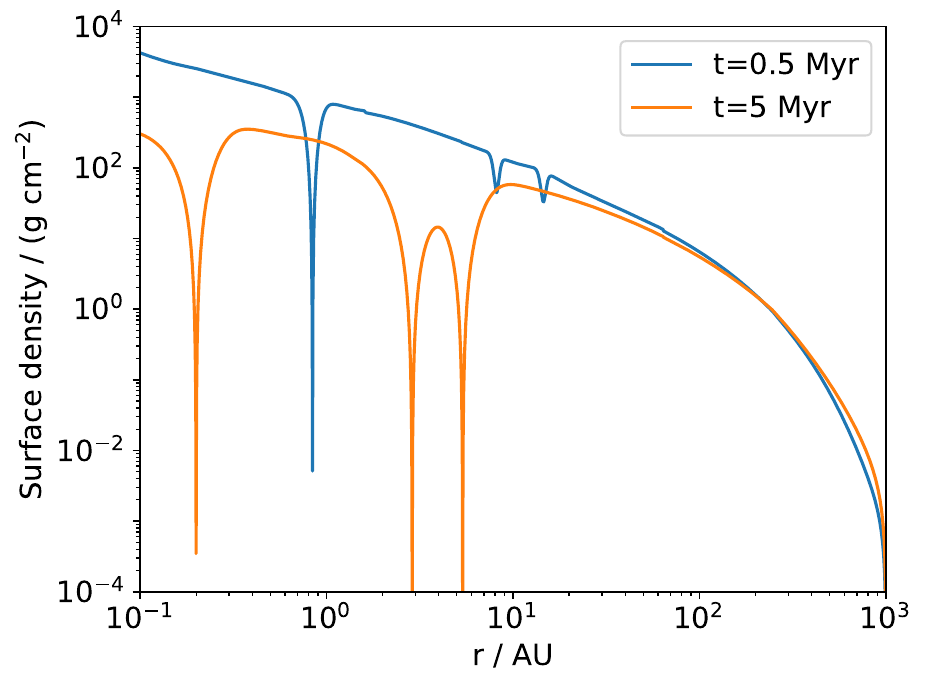}
    \caption{Gas surface density with $\alpha=5\times10^{-4}$ at time $t=0.5$~Myr~(blue) and $t=5$~Myr~(orange) including three planets (setup name V5P123).
    \label{fig:GapOverlapExample}}
\end{figure}

As the planet grows it exchanges angular momentum with the surrounding gas. Some of the momentum is carried away in the form of spiral waves, some will exert a net torque onto the planet leading to planet migration and some is deposited locally in the disk opening a gap. The gap opening by gravity can be modelled using the simple empirical model that links a parameter $\mathcal{P}$ (see eq. (25) in \citet{schneider_how_2021-1}), describing the local gas properties related to the planet, to a gap depth $f\low{gap}(\mathcal{P})=\Sigma\low{gap}/\Sigma\low{gas} \in \left[0,1\right]$, given by eq. (26) of \citet{schneider_how_2021-1})\citep{crida_width_2006, crida_cavity_2007}.
Following \citet{schneider_how_2021-1}, we use a Gaussian profile $\aleph(r;a\low{p},f\low{gap},\sigma)$ to modify $\alpha$. The Gaussian profile height is given by $f_\text{gap}$ at $a_\text{p}$ and a standard deviation $\sigma=r\low{HS}/(2\sqrt{2\ln(2)})$ depending on the horseshoe radius $r\low{HS}$ $=x\low{HS}a\low{p}$ with $x\low{HS}$ as given by eq. (48) in the paper by \citet{paardekooper_torque_2011}.
\begin{equation}
    \aleph(r;a\low{p},f\low{gap},\sigma)=1-(1-f\low{gap})\exp{\left(\frac{(r-a\low{p})^2}{2\sigma^2}\right)}
\end{equation}
The increase of the disk's viscosity at the planet's position opens a gap in the gas surface density by accelerating the transport of gas. This approach has also been used in other studies \citep[e.g.][]{dullemond_disk_2018, pinilla_growing_2021}. Given multiple planets, the total viscosity profile $\alpha'(r)$ including the distortion can be expressed by multiplying the individual profiles $\aleph_i(r)$.
\begin{equation}
    \alpha'(r)=\frac{\alpha}{\prod_{i\:\in\text{ planets}}\aleph_i(r)}
\end{equation}
We use the modified alpha profile $\alpha'(r)$ only for the calculation of the gas dynamics while for other processes $\alpha$ stays constant across the gap.
Fig. \ref{fig:GapOverlapExample} shows the result of the gap formation prescription in combination with our orbital resonance prescription. The gas surface density is shown for a simulation with three planets at two times. The early stage has shallow and well-separated gaps because the planets are not massive enough. With more massive planets, the gaps grow to an overlapping gap around the middle and outer planets. In this case, the outer planet migrates in resonance with the middle planet which prevents a too-close encounter.

\subsection{Migration}
The angular momentum exchange with the gas affects the planet's orbit and can lead to inward or outward migration. The code uses prescriptions for type~I and type~II migration and interpolates linearly between both cases. Type II is fully reached after opening a deep gap ($f\low{gap}<0.1$). 

For type I migration, the code includes the Lindblad-torque $\Gamma_\text{L}$, corotation-torque $\Gamma_\text{C}$ \citep{paardekooper_torque_2010, paardekooper_torque_2011}, thermal-torque $\Gamma_\text{T}$ \citep{lega_migration_2014, benitez-llambay_planet_2015, masset_coorbital_2017, baumann_influence_2020} and the dynamical-torque \citep{paardekooper_dynamical_2014,pierens_fast_2015} for gap opening planets.
We calculate the Lindblad-torque and the corotation-torque using eq. (3), (52) \& (53) of \citet{paardekooper_torque_2011}. 
The thermal-torque is given by eq. (146) of \citet{masset_coorbital_2017} and mostly affects small fast accreting planets \citep{baumann_influence_2020}.
Their sum yields the total torque without the dynamical scaling.
\begin{equation}
    \Gamma_\text{tot} = \Gamma_\text{L} + \Gamma_\text{C} + \Gamma_\text{T}
\end{equation}
The final Type I torque including the dynamical torque is then given by eq. (33) of \citet{paardekooper_dynamical_2014}.
Those models need the logarithmic gradient of the surface density $\alpha_\Sigma=-\frac{\text{d ln}\Sigma}{\text{d ln}r}$. Previously, this value was computed by linear interpolation between gap edges. However, when dealing with multiple gaps, defining the gap edges is less clear. For simplicity, we use the initial disk profile to calculate this gradient. In the beginning, the surface density profile doesn't change fast, especially at low viscosity. Given that planets open gaps relatively quickly and transition into type II migration, this approximation is adequate.

The type II migration occurs on the viscous timescale of the disk $\tau_\nu=a\low{p}^2/\nu$ \citep{lin_tidal_1986}. However, this underestimates the migration time scale for planets, that are more massive than the disk just outside the gap \citep{baruteau_planet-disk_2014}. Therefore the type II migration timescale $\tau\low{II}$ is scaled by a factor relating the planet mass to the disk mass just outside the gap.
\begin{equation}
    \tau\low{II}=\tau_\nu\times\max\left(1,\: \frac{M\low{p}}{4\pi\Sigma\low{0}a\low{p}^2}\right)
    \label{eq:typeII_migration_time}
\end{equation}
With $M\low{p}$ being the planet's mass and $\Sigma\low{0}$ and the surface density at the outer gap edge. We determine this value by a vanishing gradient in the surface density. Previously this value was estimated by interpolating the surface density between the gap edges. The previous approach leads to slightly faster migration for very high-mass planets.

Gravitational interaction can lead to orbital resonance between planets resulting in a different migration. For simplicity we don't use an N-body solver but instead model resonance by constraining the migration at certain resonant period ratios. This primarily impacts the relative migration of planets, and we will discuss how it influences our results. We only considered the first and most often observed \citep{winn_occurrence_2015} resonance case (2-1). This means the ratio of orbital frequencies is given by $\Omega\low{in}/\Omega\low{out}=2$ of an inner and outer planet. In addition to the ratio of the frequencies the resonance argument $\theta$ which is given by:
\begin{equation}
    \theta=2\,\Omega\low{out}\,t +\Omega\low{in}\,t,
\end{equation}
needs to stay constant or at least oscillate around a constant angle over time $t$ \citep{armitage_astrophysics_2013}. However, it would require at least 2D calculations to solve this problem exactly. By assuming circular Keplerian orbits and that planets resonate once they reach the 2-1 frequency ratios one can derive a condition for the semi-major axis of two planets. The frequencies are given by the Kepler frequency $\Omega\low{K}$ at any distance $r$ to a star of mass $M_\star$ which translates to a semi-major axis ratio of two planets in resonance ($\theta=0$)
\begin{align}
    \label{eq:orbital_resonance}
    0&=2a\low{out}^{-3/2} - a\low{in}^{-3/2}\\
    \Rightarrow \frac{a\low{out}}{a\low{in}}&=2^{2/3}\approx1.58740105 
\end{align}
We assume that once a planet crosses the resonant semi-major axis of another planet, both are captured in resonance. They remain in resonance and migrate according to the inner planet's migration rate. While this approach is a simplification it serves our purpose of studying how multiple planets influence the disk and atmosphere composition.

\subsection{Initial conditions}
\begin{table}[htb]
    \caption{Initial setup overview}
    \label{tab:simulation_setups}
    \centering
    \small
    \begin{tabular}{|c|c|c|c|c|}
        \hline
        Setup  & Viscosity $\alpha$ & Planet 1 & Planet 2 & Planet 3\\\hline
        V1noP & $1\times10^{-4}$ & & & \\
        V1P1 & $1\times10^{-4}$ & \checkmark & & \\
        V1P2 & $1\times10^{-4}$ & & \checkmark & \\
        V1P3 & $1\times10^{-4}$ & & &  \checkmark \\
        V1P12 & $1\times10^{-4}$ & \checkmark & \checkmark & \\
        V1P13 & $1\times10^{-4}$ & \checkmark & & \checkmark \\
        V1P23 & $1\times10^{-4}$ & & \checkmark & \checkmark\\
        V1P123 & $1\times10^{-4}$ & \checkmark & \checkmark & \checkmark \\
        \hline
        V5noP & $5\times10^{-4}$ & & & \\
        V5P1 & $5\times10^{-4}$ & \checkmark & & \\
        V5P2 & $5\times10^{-4}$ & & \checkmark & \\
        V5P3 & $5\times10^{-4}$ & & &  \checkmark \\
        V5P12 & $5\times10^{-4}$ & \checkmark & \checkmark & \\
        V5P13 & $5\times10^{-4}$ & \checkmark & & \checkmark \\
        V5P23 & $5\times10^{-4}$ & & \checkmark & \checkmark\\
        V5P123 & $5\times10^{-4}$ & \checkmark & \checkmark & \checkmark \\
        \hline
        V10noP & $10\times10^{-4}$ & & & \\
        V10P1 & $10\times10^{-4}$ & \checkmark & & \\
        V10P2 & $10\times10^{-4}$ & & \checkmark & \\
        V10P3 & $10\times10^{-4}$ & & &  \checkmark \\
        V10P12 & $10\times10^{-4}$ & \checkmark & \checkmark & \\
        V10P13 & $10\times10^{-4}$ & \checkmark & & \checkmark \\
        V10P23 & $10\times10^{-4}$ & & \checkmark & \checkmark\\
        V10P123 & $10\times10^{-4}$ & \checkmark & \checkmark & \checkmark\\
        \hline
    \end{tabular}
    \tablefoot{Planet 1 starts at 3 AU, Planet 2 at 20 AU and Planet 3 at 50 AU. The numbers after the letter ``P'' indicate which planets are present in a setup. We call the viscosities with $\alpha=1\times10^{-4},\,5\times10^{-4}\,\&\, 1\times10^{-5}$ low, mid and high viscosity respectively. The number after the letter ``V'' indicates the viscosity. See Tab. (\ref{tab:default_setup}) for parameters, that don't change between setups.}
\end{table}
\begin{table}[htb]
    \caption{Simulation parameters, that don't change between setups.}
    \label{tab:default_setup}
    \centering
    \small
    \begin{tabular}{|c|c|c|}
        \hline
        Parameter       & Value                 & Description\\
            \hline\hline\multicolumn{3}{|c|}{Disk}\\\hline
        $M_\star$       & $1.0\,\text{M}_\odot$          & mass of central star\\
        $L_\star$       & $1.0\,\text{L}_\odot$        & luminosity of central star\\
        $\alpha\low{z}$ & $1\times10^{-4}$      & vertical stirring parameter\\
        $M_0$           & $0.128\,\text{M}_\odot$        & initial disk mass\\
        $R_0$           & $137\,\text{AU}$        & disk scaling radius\\
        $\epsilon_0$    & $0.0124$              & initial solid to gas ratio\\
        $v\low{f}$      & $5.0$ m/s             & fragmentation velocity \\
        $t\low{evap}$   & $10.0\,\text{Myr}$       & disk lifetime before dispersal\\
        $\tau\low{decay}$& $0.01\,\text{Myr}$     & dispersal time scale\\
            \hline\hline\multicolumn{3}{|c|}{Planet}\\\hline
        $t_0$           & $0.05\,\text{Myr}$      & planet starting time\\
        $\rho\low{c}$         & $5.5\, \text{g / cm}^3$    & density of the planetary core\\
            \hline\hline\multicolumn{3}{|c|}{Simulation}\\\hline
        $r\low{in}$     & $0.1\,\text{AU}$        & inner edge\\
        $r\low{out}$    & $1000\,\text{AU}$       & outer edge\\
        $N\low{Grid}$   & $5000$                 & number of grid cells\\
        $\Delta t$      & $10\,\text{yr}$         & integration time step\\
        \hline
    \end{tabular}
\end{table}

To study the influence of multiple planets on the inner disk's water abundance we prepare a set of simulations that vary in the number of planetary seeds (from zero to three) and we consider three different disk viscosities with $\alpha=1,\, 5\,\&\, 10\times10^{-4}$, which we will call low, mid and high viscosity respectively. The planets start on one out of three initial positions with $a\low{p,0}=3,\,20\,\&\,50$ AU. We call the planet that starts at 3 AU Planet 1, the planet starting at 20 AU Planet 2 and the planet starting at 50 AU Planet 3. Tab. (\ref{tab:simulation_setups}) gives an overview of the different simulations and Tab. (\ref{tab:default_setup}) lists parameters that don't change between the setups.

\section{Results}\label{sec:results}
\subsection{Disk water content}
\begin{figure*}[htb!]
    \centering
    \includegraphics[width=0.99\linewidth]{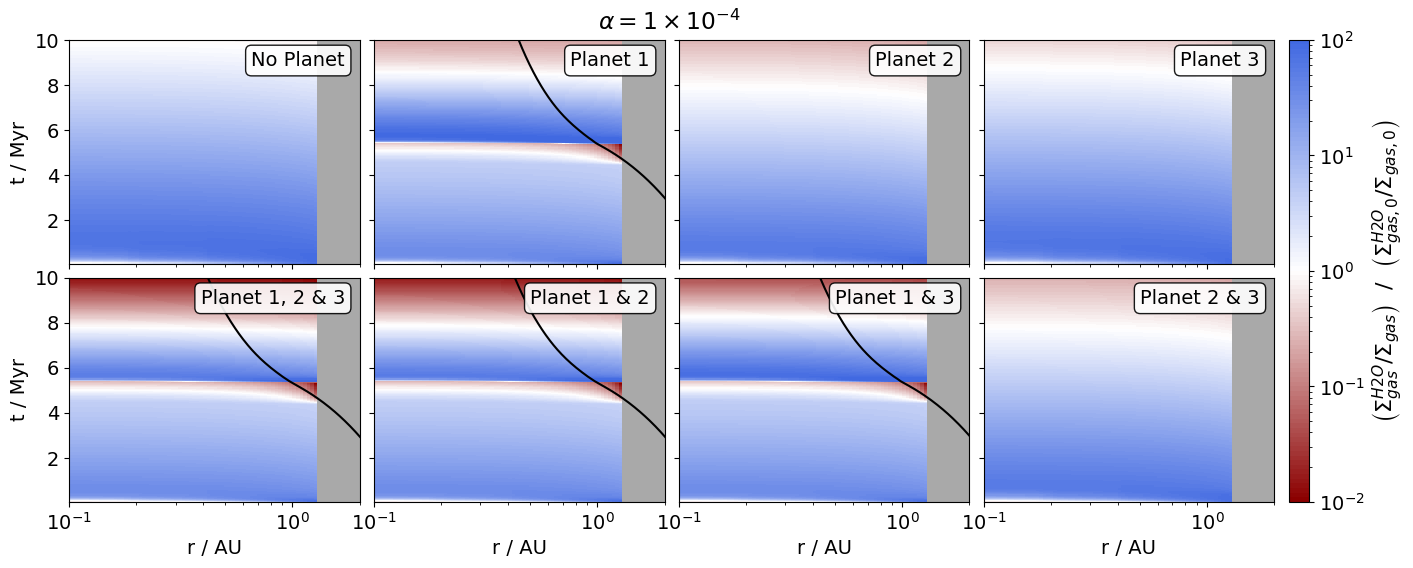}
    \vspace{-3pt}
    \includegraphics[width=0.99\linewidth]{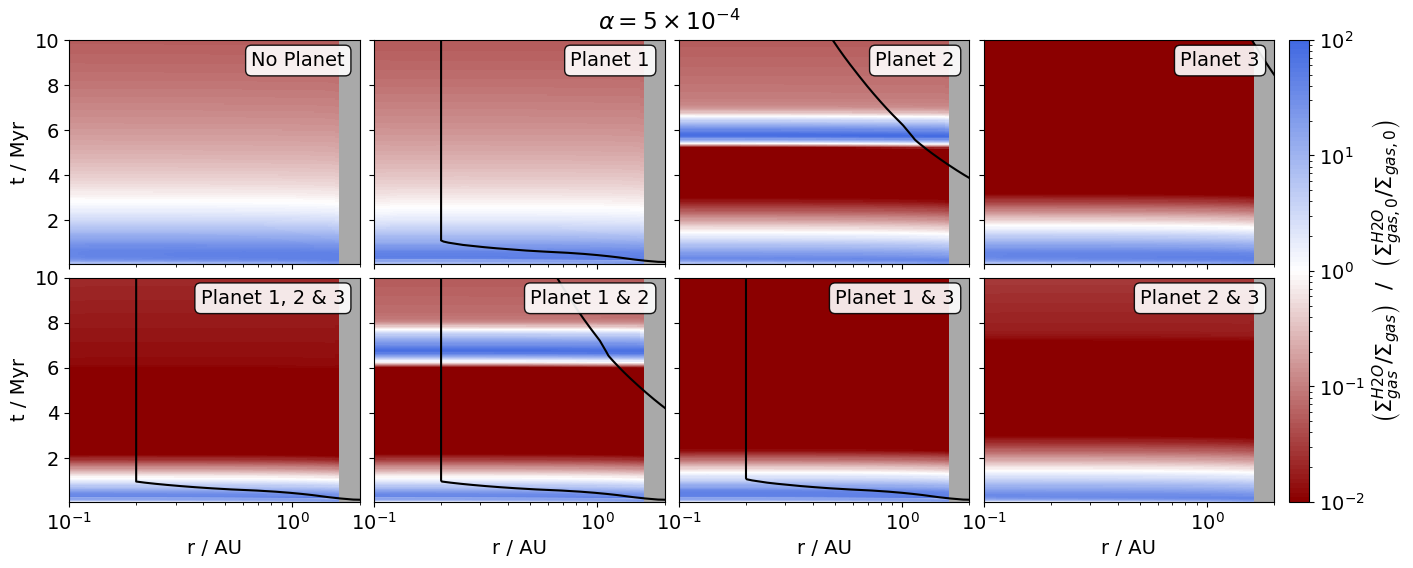}
    \vspace{-3pt}
    \includegraphics[width=0.99\linewidth]{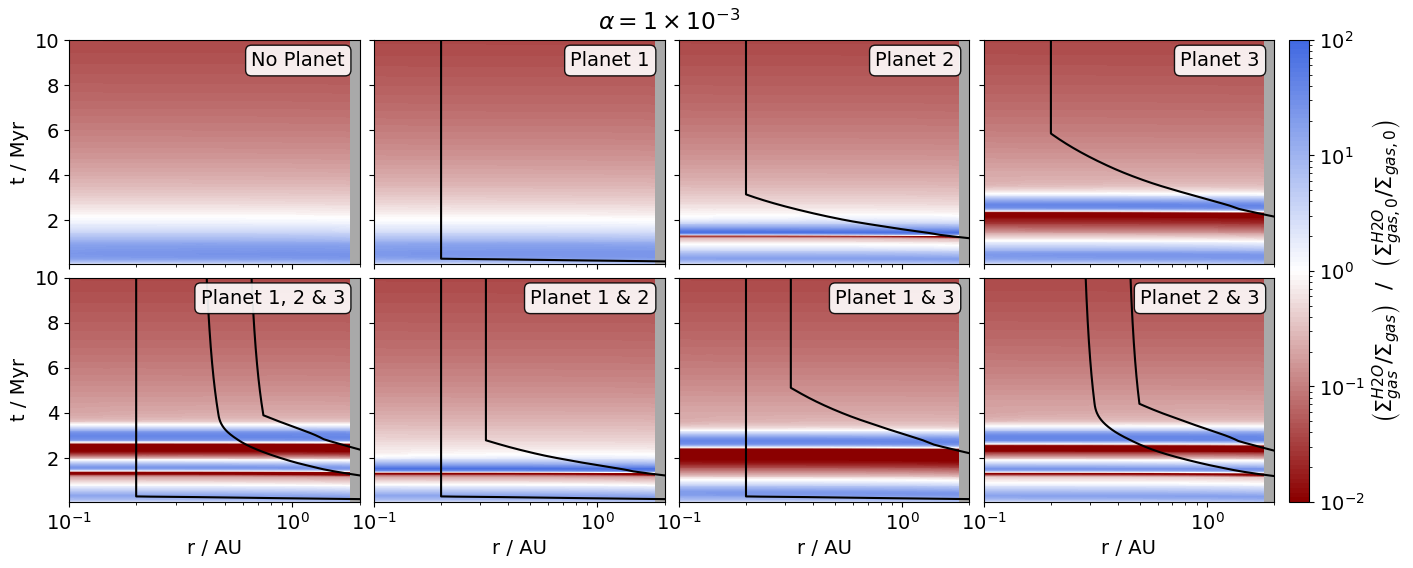}
    \caption{Water fraction in the gas surface density normalised to the initial fraction for $\alpha=1,\, 5,\,\&\, 10\times10^{-4}$ from top to bottom. The black lines represent the migration paths of planets. The grey area shades the region exterior to the water ice line. Note, only some plots show the path of planets because depending on the setup either no planets are present or they remain outside of 2 AU.}
    \label{fig:WaterHeatmapAll}
\end{figure*}
First, we study the water vapour fraction inside the water ice line. Fig. (\ref{fig:WaterHeatmapAll}) shows the water surface density in the gas phase relative to the total gas surface density as a function of the radius (x-axis) and the time (y-axis) normalised by the initial fraction.
For every viscosity, we show the case of no planet, one, two and three planets and every combination of two planets. 
The simulations without planets are consistent with the results of \citet{schneider_how_2021-1}. The water content depends on the delivery of icy pebbles and the viscous gas transport. 

The evaporation of volatile-rich pebbles causes a density enhancement at evaporation fronts, leading to enhanced inward diffusion of material compared to a purely viscous evolution. Water vapour is transported on a viscous timescale when the water fraction is distributed uniformly.
Because we solve eq.~(\ref{eq:viscous-evolution}) independently for every species, we find the water vapour velocity to be slower than the H \& He gas by a factor of $\sim1/2$ at low viscosity after $\sim 1$ Myr. Further discussion on the velocity and transport of the water can be found in appendix \ref{sec:WaterTransport}.

At low viscosity, just outside the water ice line, the particle size is limited by fragmentation for $\sim1$ Myr (see Fig. (\ref{fig:StokesNumber})). In this case due to the larger pebble sizes the transport of water-rich solids is very efficient. This increases the pebble flux across the ice line. In comparison, in the case of low viscosity the gas transport is slow. Therefore, the high water fraction in the gas is retained for a longer time.

Increasing the $\alpha$-viscosity value affects the viscous transport of gas, as well as the drag and size of pebbles. 
Pebbles are generally smaller and slower at higher viscosity resulting in a lower ice transport towards the ice line. 
The viscous transport is more efficient and moves water-rich gas faster towards the star. Therefore the excess in the water fraction (compared to the initial fraction) does not persist for more than $t\sim2$ Myr in the high viscosity case.

Placing Planet 1 starting at 3 AU (black line in Fig. \ref{fig:WaterHeatmapAll}) in the low viscosity disk shows how the water enrichment is lower before the planet crosses the ice line. Especially just before the crossing, the water fraction decreases below the initial value, because the planet blocks water-rich pebbles once it reaches a sufficient mass to open up a deep gap. However, the pebbles pile up outside of the planet-induced gap and as soon as the planet migrates over the water ice line, such that the outer gap edge of the planet is inside the water ice line, all the water-rich pebbles can evaporate relatively quickly. This causes the water fraction to increase to very high values in a relatively short period. After this burst of water vapour, the water vapour is carried away by viscous transport, resulting in a water-poor inner disk a few Myr after the planet has crossed the water ice line. Planet 2 \& Planet 3 do not cross the water ice line and block pebbles further outside, lowering the water fraction below the initial value. This effect is smaller when the planet is located further away from the ice line.

For the cases with Planet 1 and at least one other planet the late evolution of the disk from $\sim 8$ Myr onwards is sensitive to a second planet that does not cross the ice line. Its presence reduces the water fraction at the end of the disk's lifetime by an order of magnitude. With a smaller orbit of the second planet (i.e. having Planet 2 instead of Planet 3) the enrichment is more reduced. However, the cases of having only Planet 2 and Planet 2 \& Planet 3 are very similar.

As the viscosity increases, the planets can migrate faster. The effect of Planet 1 on the water fraction for the mid and high-viscosity disk is very small. It crosses the ice line early such that it can neither block the pebbles with a gap nor accrete a lot of water-rich pebbles. 

However, at mid-viscosity, Planet 2 blocks the pebbles before crossing the ice line around 6 Myr. We thus observe the same pattern as for the innermost planet (Planet 1) in the low viscosity case: the inner disk experiences an outburst of water vapour due to the evaporation of pebbles that are trapped by the migrating planet. Planet 3, in this case, does not cross the water ice line and consequently only blocks the water-rich pebbles, clearly to sub-stellar water abundance in the disk after 2 Myr.

In the case of Planet 1 in combination with Planet 2, we find that the migration of Planet 2 is affected by Planet 1 because the surface density is lower when two planets are present in the disk lowering the value of $\Sigma\low{0}$ for Planet 2. This increases the migration time scale as given by eq.~(\ref{eq:typeII_migration_time}) compared to the case when only Planet 2 is present.
In the setup that includes Planets 2 \& Planet 3 the migration of both planets is slower as they open up an overlapping gap and migrate in resonance (not shown in Fig.~\ref{fig:WaterHeatmapAll}). In our prescription, this leads to slower migration because the surface density just outside of Planet 2's gap is significantly lower. Therefore there is no late enrichment with water as Planet 2 never crosses the ice line.
With all three planets, the water content in the inner disk rapidly decreases to very low values due to the efficient blocking of water-rich pebbles in the outer disk.

For the high viscosity case, all planets migrate enough to cross the water ice line.
This leads to multiple short periods of water enrichment due to the evaporation of blocked pebbles. Except with only Planet 1, the enrichment follows the pattern without a planet because Planet 1 crosses the water ice line during the early pebble enrichment phase.
The overall depletion of water inside the water ice line is delayed until the last planet crosses the water ice line. In our setups, this prolongs the time the disk contains high fractions of water to about 3.5 Myr for setups including Planet 3 compared to 2 Myr for the disk-only setup. However, all the setups have similar water fractions after about 4 Myr i.e. after the last planet crossed the ice line and trapped pebbles evaporated.

\begin{figure}[ht!]
    \includegraphics[width=0.99\linewidth]{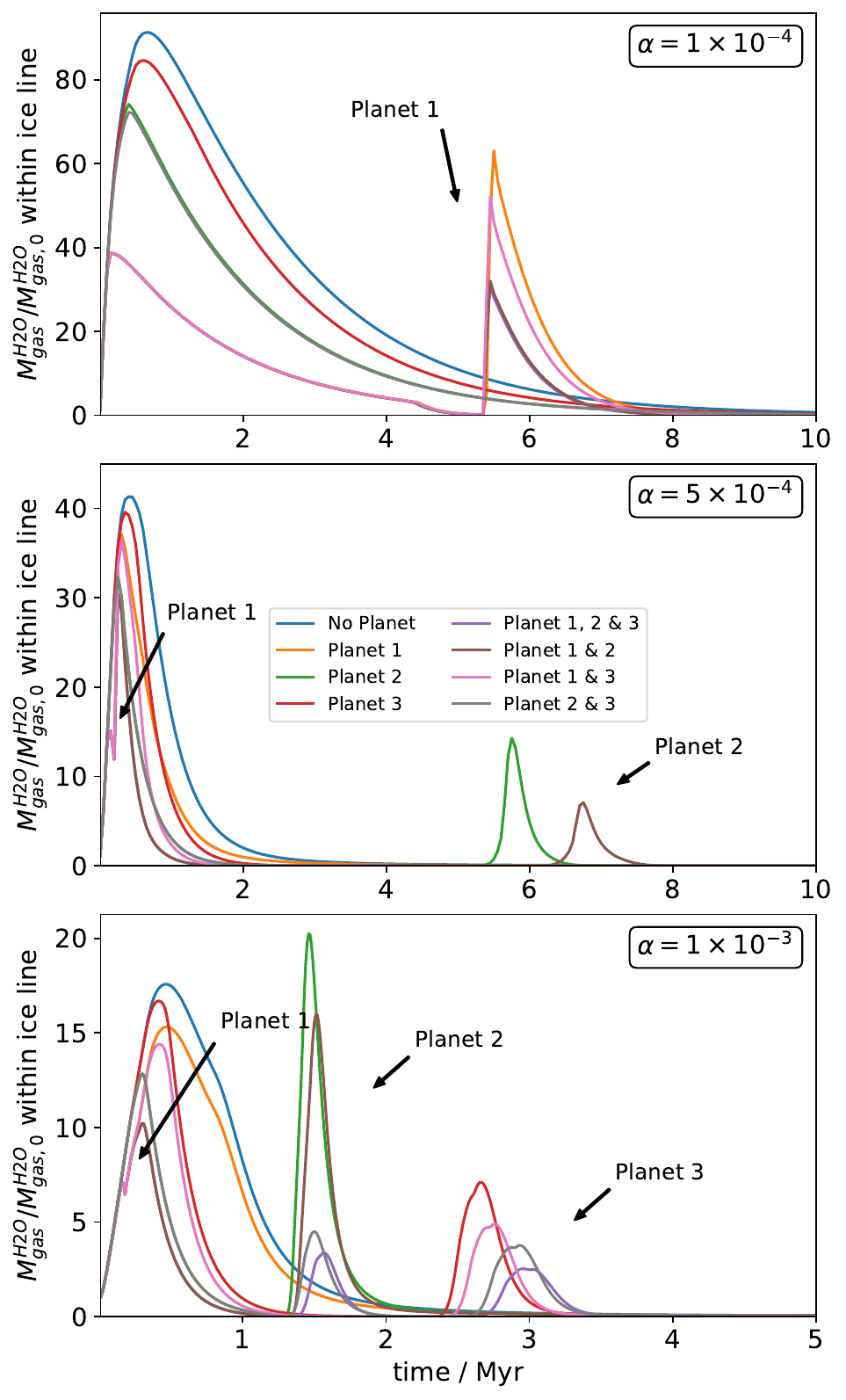}
    \caption{Water mass fraction within the ice line (i.e. integrated over the region with a radius smaller or equal to the position of the water ice line) normalised by the initial value against time. The panels from top to bottom show the three disk viscosities. Note for $\alpha=1\times10^{-3}$ the x-axis only spans to $t=5$ Myr, but the late evolution doesn't deviate from 0 which can also be seen in Fig. (\ref{fig:WaterHeatmapAll}). The colours correspond to different simulation cases and arrows indicate which planet causes the water vapour to increase by crossing the water ice line.
    \label{fig:WaterInIceLineAll}}
\end{figure}

Fig. (\ref{fig:WaterInIceLineAll}) shows the total mass fraction of water vapour within the ice line normalised by the initial value.
Note the y-axis has different limits for the three different viscosities and the highest enrichment can be observed for the lowest viscosity. The amount of water vapour enrichment depends on the pebble flux, their size and gas drift velocity as described above. Again we observe the initial phase of enrichment (within the first 1-4 Myr depending on the viscosity) where planets are relatively small, and the phase of water enrichment every time a planet crosses the water ice line. The disk with no planets shows the largest initial water content because no planets are present that can reduce the inward flux of pebbles. Depending on the distance of a planet to the ice line the enrichment is reduced. The closest planet can reduce the water effect by 50\% while the most distant planet has only a small effect. 

Focusing on the low viscosity, there is almost no difference between the cases of e.g. Planet 1 alone and Planet 1 \& Planet 2. The innermost planet has always the strongest effect. However, for the second peak, which is caused by Planet 1 crossing the ice line there is a clear distinction between no other planet and a second or third planet. The other planets affect the available solid material that can evaporate at this later secondary phase. In the configuration involving only Planet 1, the enrichment is about 70 times above the initial water vapour mass while for the configuration with Planet 1 \& Planet 3 it is $\sim 60$ and for Planet 1 \& Planet 2 $\sim40$. Note the configuration with all planets does not deviate from the configuration with only Planet 1 \& Planet 2. Hence a second planet can reduce the ice line crossing-induced water enrichment of the gas by a factor of up to 50\%.

For the mid- and high-viscosity, not only the peak amount of water mass is reduced but also the differences between the setups decrease. Nevertheless, the secondary peaks caused by planets crossing the ice line remain more sensitive to the presence of outer planets. Also, Planet 1 migrates relatively quickly inside the water ice line such that its effect on the water mass inside the ice line is small and mostly changes the shape of the initial enrichment peak. 
In the high-viscosity case, the initial peak of water enrichment is the smallest, as discussed above. Interestingly, once the first planet starts to migrate interior of the water ice line, the peak of water enrichment is similar to the initial peak without any planet.

In principle, the effect of gaps depends on many parameters such as gap location, depth and viscosity, which in our case depend on the evolving planet properties. Nevertheless, our results in Fig. (\ref{fig:WaterInIceLineAll}) are similar to previous studies \citep{kalyaan_effect_2023}. Namely, the amount of enrichment strongly depends on the innermost gap (or planet in our case) outside the water ice line. The main difference to the previous study originates from the gap evolution, which in our case depends on the changing planet properties.

\subsection{Disk C/O ratio}
\begin{figure*}[htb!]
    \centering
    \includegraphics[width=0.99\linewidth]{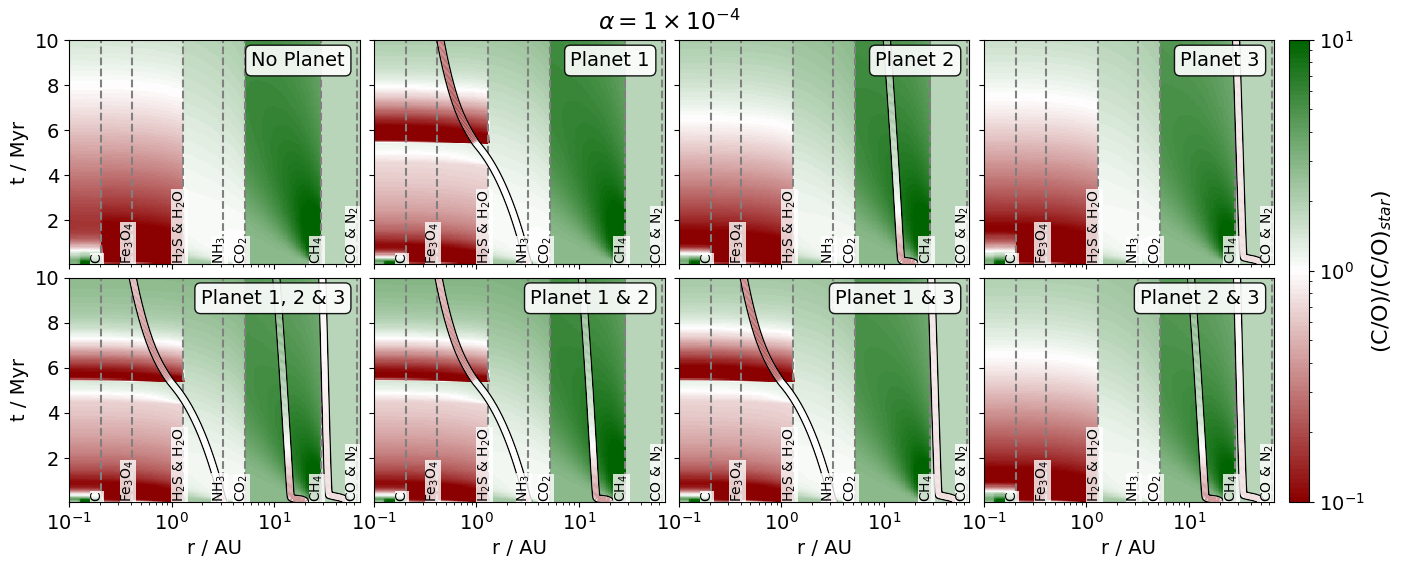}
    \includegraphics[width=0.99\linewidth]{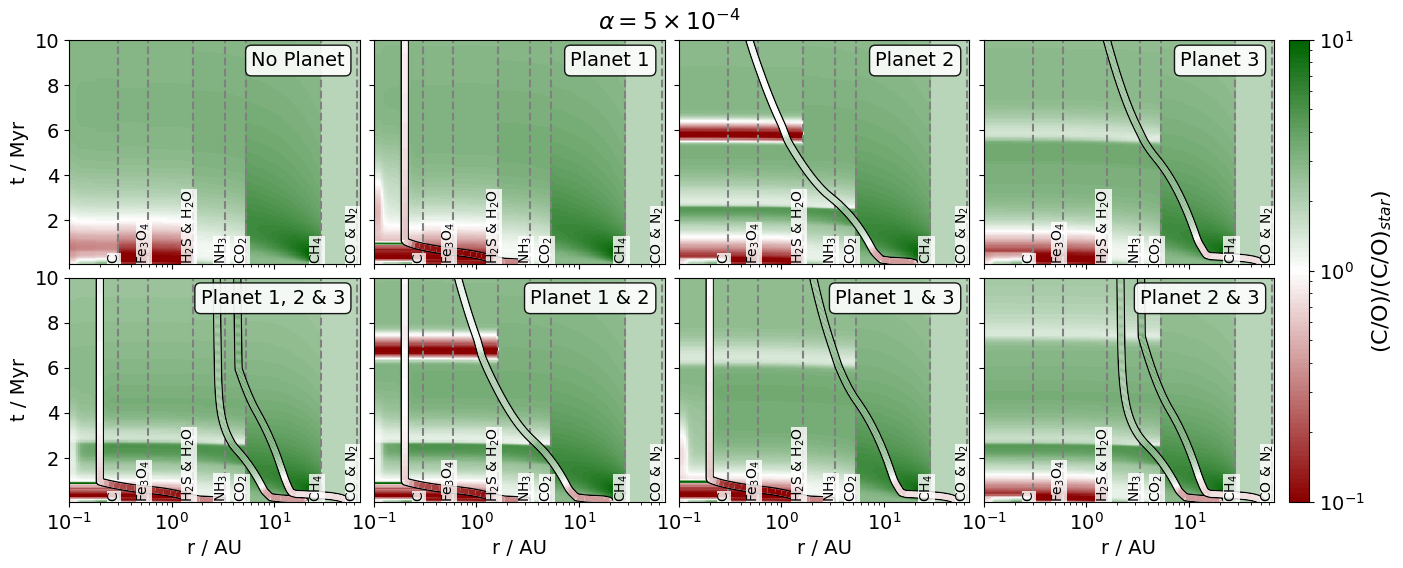}
    \includegraphics[width=0.99\linewidth]{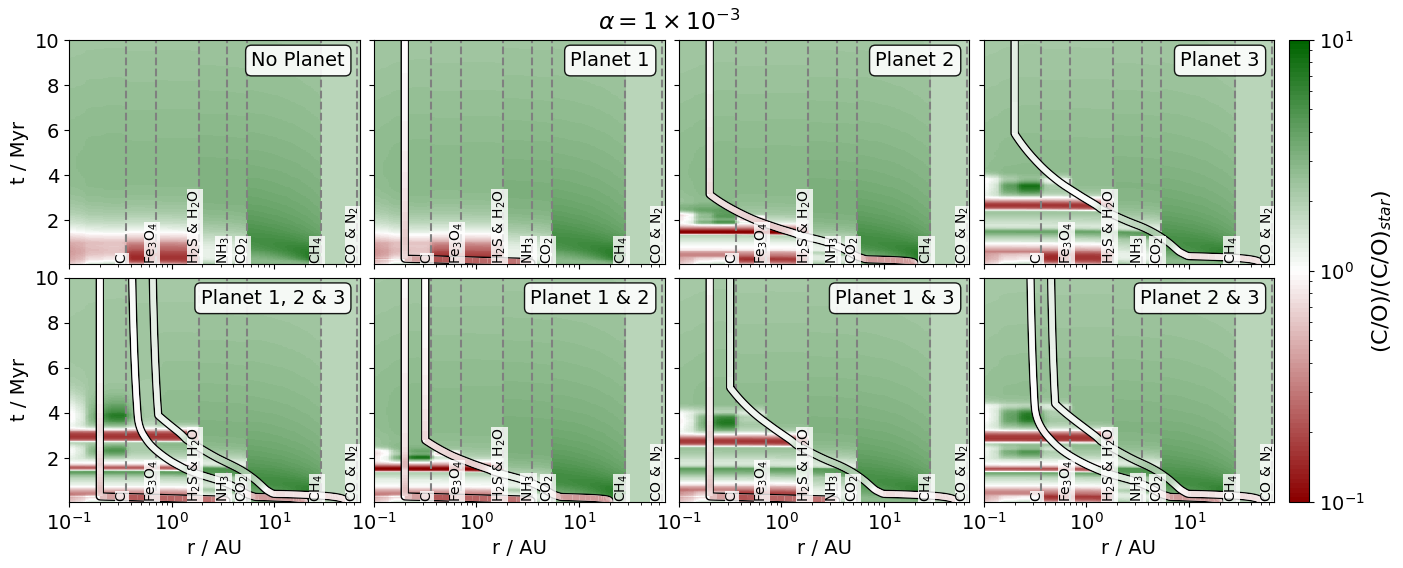}
    \caption{C/O number ratio in the gas phase normalised by the stellar value. The panels from top to bottom show the different disk viscosities $\alpha=1,\, 5,\,\&\, 10\times10^{-4}$. The coloured line (enclosed between two black lines for visibility reasons) shows the position of the planet and the current bulk C/O number ratio in the atmosphere normalised by the stellar value.\label{fig:COHeatmapAll}}
\end{figure*}
In this section, we investigate how the disk's C/O ratio is affected by pebble drift and evaporation in the presence of multiple growing planets. Fig. (\ref{fig:COHeatmapAll}) shows the C/O number ratio normalised to the stellar value. Additionally, the bulk C/O ratio of the atmosphere of the growing planet(s) is shown with a thick line enclosed by two black lines. The colour scale representing the C/O ratio is the same for the disk and the planet.

The most significant oxygen enrichment compared to carbon originates from water, explaining the similarities in the inner region to Fig. \ref{fig:WaterHeatmapAll}. The highest ratio of carbon is reached between the evaporation fronts of CO$_2$ and CH$_4$. The available solid material from CH$_4$ is 25\% of all carbon species (see Tab. (\ref{tab:mixing_ratios})), which can significantly increase the C/O ratio while CO$_2$ decreases it again. In our chemical model, CO$_2$ only makes up 10\% of all carbon species. 
Planets that accumulate substantial gas from the region between the ice lines of CH$_4$ and CO$_2$ can increase their atmospheric C/O ratio.
However, blocking pebbles in this region doesn't change the C/O ratio in the disk.

The plot shows how the atmosphere C/O changes when a planet crosses an evaporation front. For example, at low viscosity, the atmospheric C/O ratio rapidly decreases after Planet 1 crosses the water ice line. However, for Planet 2 the increase of the C/O ratio in its atmosphere occurs on a much longer timescale despite staying in a high C/O ratio part of the disk. For Planet 1 which ends up having a sub-stellar C/O  ratio the effect of an outer planet is observable due to the blockage of water-rich pebbles. 

We note that the final C/O ratio in the disk depends on the initial chemical partitioning model. For example, lower methane fractions will reduce the C/O ratio between the CO$_2$ and CH$_4$ evaporation fronts and thus also change planetary compositions \citep[e.g.][]{schneider_how_2021}.

\subsection{Bulk atmosphere composition}
\begin{figure}[!htb]
    \includegraphics[width=0.99\linewidth]{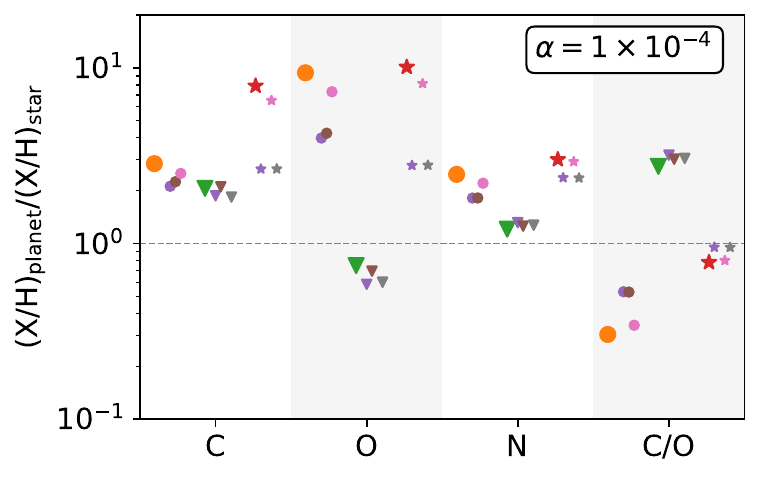}
    \includegraphics[width=0.99\linewidth]{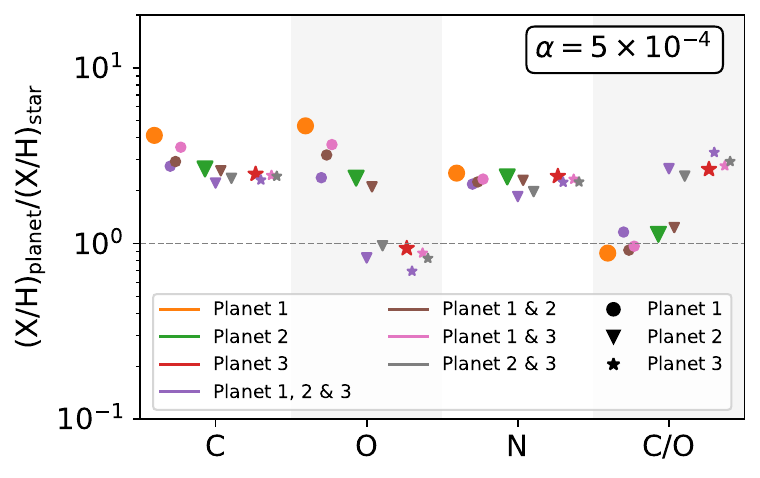}
    \includegraphics[width=0.99\linewidth]{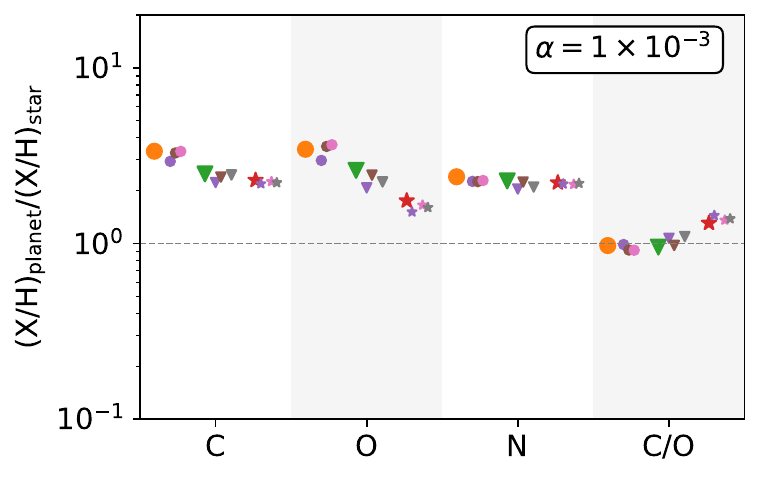}
    \caption{Chemical abundance in the atmosphere at t=10 Myr and different disk viscosities $\alpha=1,\, 5\,\&\, 10\times10^{-4}$ (top to bottom). The different colours show the different configurations and the different markers show the given planet. The markers for the single planet cases are larger. For the elements C, O, and N the abundances are given in the number ratio to hydrogen normalised by the stellar value while C/O is computed using (C/H)/(O/H) normalised by the stellar value. For the exact numbers see Tab. (\ref{tab:Simulation_Results}).\label{fig:ChemChomp}}
\end{figure}

Below we investigate if the bulk atmospheric composition of C, O, N and the C/O ratio is altered when multiple planets are present. Fig. (\ref{fig:ChemChomp}) shows the final abundances of those elements (relative to hydrogen) normalised by the solar value. Planet 1 is represented by a dot, Planet 2 by a triangle and Planet 3 by a star. To make it comparable with Fig. (\ref{fig:WaterInIceLineAll}) we used the same colours for the same setups.
In the following, we compare the abundances of the multi-planet cases to the single-planet case. The exact numbers can be found in Tab. (\ref{tab:Simulation_Results}) in the appendix.

\subsubsection{Low-viscosity}
In the low viscosity case, all planets have super-solar carbon abundances. 
The carbon abundance in the atmosphere of Planet 1 is reduced by $\sim20\%$ if Planet 2 is present due to the blockage of CO$_2$ and CH$_4$ rich pebbles. Planet 3 decreases the carbon abundance in the atmosphere of Planet 1 by $\sim10\%$. If all planets are present, the carbon abundance of the atmosphere of Planet 1 is reduced by $\sim25\%$.

The carbon abundance of Planet 3 can not be explained by the blocking of pebbles because no other planet could alter the chemical composition of the surrounding gas of Planet 3. However, Planet 3 can grow more massive when Planet 2 is present, because of a higher disk accretion rate, which limits the late gas accretion of Planet 3.
In addition, the opening of the gap by Planet 2 pushes more gas towards Planet 3. Similar effects have been observed in 2D hydrodynamic simulations \citep{bergez-casalou_simultaneous_2023}, where an outer planet can accrete gas slightly more efficiently with an inner companion, especially at lower viscosities. Hence Planet 3's carbon-rich primordial atmosphere is more diluted with low carbon-enriched gas when Planet 2 is present (see $M_\text{env}/M$ in Tab. (\ref{tab:Simulation_Results})).

The oxygen abundance in the atmosphere of Planet 1 strongly depends on the amount of accreted water. In the presence of Planet 2 (Planet 3), the abundance of oxygen in Planet~1 is reduced by $\sim 55\%$ ($\sim20\%$). The oxygen reduction is primarily controlled by Planet 2, and having more planets in the disk has a negligible effect on the carbon abundance. This, unsurprisingly, indicates that the closer planet is more important for the evolution of the final atmospheric abundances than planets further away.
 
The sub-solar oxygen abundance in the atmosphere of Planet 2 is caused by its position relative to the oxygen-bearing molecules. Planet 2 is far away from the CO evaporation front but also does not cross the CO$_2$ evaporation front. We find that pebble blocking by Planet 3 changes the oxygen abundance in the atmosphere of Planet~2 by $\sim20\%$. 
The difference in the inferred atmospheric composition of Planet 3 is not related to blocked pebbles but to a higher atmosphere mass as explained above.
 
The abundance of nitrogen in the atmosphere of Planet 1  depends on the blocking of NH$_3$ rich pebbles. Planet 2 and Planet 3 reduce the nitrogen abundance in the atmosphere of Planet 1 by about $\sim 25\%$ and $\sim 10\%$, respectively. The N$_2$ containing pebbles are located far out in the disk where the viscous timescale is too long to transport N$_2$ enriched gas to planets closer to the star (see Fig. (\ref{fig:DiskNitrogenAbundance})). 

Planet 3's atmosphere built during the pebble accretion phase has a low nitrogen abundance because only small amounts of NH$_3$-rich pebbles were accreted. As soon as the planet accretes gas, which is heavily enriched by N$_2$, the bulk value for N/H approaches that of the surrounding disk independent of the atmosphere mass. Note that in every simulation N/H is found to be super-stellar due to the very volatile nature of the nitrogen components. Similar to O/H, the N/H ratio of Planet 2 is slightly lower compared to Planet 1 and Planet 3, due to its special position: far interior of the N$_2$ evaporation front, but exterior to the NH$_3$ evaporation front.

The C/O ratio in the atmosphere of Planet 1 is governed by the accretion of water and CO$_2$. These species are blocked by Planet 2 such that the C/O ratio in the atmosphere of Planet 1 is increased by $\sim 75\%$. Although Planet 3 can block the same species and additionally some methane, Planet 2 dominates the C/O ratio change. The C/O ratio in the atmosphere of Planet 2 is increased by $\sim10\%$ when Planet 1 or Planet 3 is present, and by $\sim15\%$ when both planets are included.  Some of this change is linked to the difference in the envelope mass fraction. 
The C/O ratio in the atmosphere of Planet 3 depends on both, the ratio in the solids as well as in the gas. 
A significant mass of the planet's envelope is a result of pebble accretion. 
Our simulations indicate that at low viscosity, super-stellar C/O ratios typically correspond to the middle planet, which is located in a region where water is in icy form, but most carbon-bearing species are already evaporated. This allows the planet to accrete gas with a high C/O ratio, see Fig. (\ref{fig:COHeatmapAll})

\subsubsection{Mid-viscosity}
For the mid-viscosity case, we have similar trends as in the low-viscosity case for the atmosphere of Planet 1. The carbon abundance is reduced by $\sim30\%$ when Planet 2 is present, by $\sim15\%$ if Planet 3 is present, and by $\sim35\%$ when both planets are present. The carbon abundance in the atmosphere of Planet 2 changes slightly because of deviations in the migration. We find that for Planet 3 the carbon abundance in the atmosphere is unchanged. 

For Planet 1 the oxygen abundance in the atmosphere strongly depends on the presence of other planets as they can reduce oxygen by $\sim50\%$ when all the three planets are present. When only one planet is present the reduction is of the order of $\sim30\%$ (Planet 2) and $\sim20\%$ (Planet 3). Blocking oxygen-rich ice pebbles depends on both outer planets because the presence of Planet 3 determines whether Planet 2 crosses the water ice line (see Fig. (\ref{fig:COHeatmapAll})). This explains the large difference in the oxygen abundance in the atmosphere of Planet 2 with a reduction of $\sim65\%$ in the presence of Planet 3. As we discussed earlier, this reduction is not due to blocked pebbles.

Nitrogen is reduced in the atmosphere of Planet 1, Planet 2 and Planet 3 by $\sim15\%$, $\sim25\%$ and $\sim10\%$, respectively. The reduction is most significant in the atmosphere of Planet 2 because Planet 3 can block NH$_3$ rich pebbles. However, in all the cases the atmospheric nitrogen abundance is found to be super-stellar.

The C/O ratio in the atmosphere of Planet 1 is very high with an increase of $\sim32\%$ if all planets are present. As discussed previously, this can be caused by the blocking of water-rich pebbles. Planet 2 and Planet 3 individually increase the C/O ratio in the atmosphere of Planet 1 by only $\sim10\%$. The biggest change in the C/O ratio with an increase by more than $100\%$ can be seen in the atmosphere of Planet 2 because Planet 2 does not cross the water ice line if Planet 3 is present. Unlike in the low viscosity case, Planet 3 can also have a large super-stellar C/O ratio in its atmosphere if it migrates inward without crossing the water ice line. This suggests that the migration history of planets has a crucial effect on the final atmospheric composition.

\subsubsection{High-viscosity}
The migration timescale strongly depends on the viscosity, especially in the late evolution when migration matches the viscous time scale. The migration paths slightly deviate between the setups but all planets migrate relatively fast inwards and line up according to the orbital resonance given by eq. (\ref{eq:orbital_resonance}). 
Therefore the blocking of pebbles is most important for the innermost evaporation lines namely of the carbon grains, iron oxides (relatively low abundance) and water. 
The atmosphere of Planet 1 is less enriched by $\sim15\%$ in carbon and oxygen. In the atmosphere of Planet 2, the carbon and oxygen  abundances are reduced by up to $\sim10\%$ and $\sim20\%$, respectively. The deviations in the atmosphere of Planet 3 are less than $\sim5\%$ for carbon and less than $\sim15\%$ for oxygen. The nitrogen abundance in the atmosphere is consistently super-stellar and changes less than $\sim 5\%$ for Planet 1, less than $\sim20\%$ for Planet 2, and less than $\sim5\%$ for Planet 3. For the C/O ratio in the atmosphere, the deviations are lower than $\sim5\%$ for Planet 1, $\sim15\%$ for Planet 2 and $\sim10\%$ for Planet 3.

\section{Caveats}\label{sec:ModelLimitations}
Our model is rather simple since it is based on modelling disk dynamics using a 1D semi-analytical prescription and a simple two-layer planet model. Therefore, some of our results may change when considering more sophisticated simulations. 

First, the planet structure is assumed to be a simple core+envelope and therefore claims in our study about the bulk composition may not be observable in planets that have more complex structures such as composition gradients \citep[e.g.][]{vazan_explaining_2020} or an interplay between the atmosphere and interiors. In addition, in reality, the planetary bulk composition could change after formation via several processes such as atmospheric escape and giant impacts \citep[e.g.][]{genda_hydrodynamic_2004, guillot_giant_2022}.
Atmospheric escape is not expected to play an important role during the formation phase. When the planet is small, the atmosphere is coupled to the disk and the upper layers of the atmosphere should have similar composition to the disk gas. Also during the detached phase, when the planet is massive enough to open a gap,  atmospheric escape is not very likely. This is because when the planet is still surrounded by the disk, the disk shields the atmosphere from high energy radiation of the star. 
Furthermore, we assume that 10\% of the accreted pebbles contribute to the initial atmosphere rather than the core during core formation. This value affects the chemical abundances in a giant atmosphere and it was shown that this ratio strongly depends on the local formation conditions \citep{valletta_giant_2020}.  However, in our case, this will mostly change the results for Planet 3 when it grows to different masses depending on the presence of Planet 2.

Second, we do not calculate the gravitational forces between planets which would affect the mean-motion resonance. Using an N-body integrator would properly account for such interactions and in addition include other mean-motion-resonance frequencies resulting in different spacing of planet pairs and resonant chains. Also, depending on the migration speed, resonant period ratios can be crossed and scattering could occur \citep{bitsch_eccentricity_2020, bitsch_giants_2023}. 
However, this will affect our results only in the case of multiple planets because, in the case of a single planet, the migration rate is determined solely by the planet-disk interaction.  
With multiple planets the different resonant ratios can lead to different configurations, however, the key trends inferred in this work regarding the enrichment are expected to remain. 

Third, our simplified gap formation method naturally extends to overlapping gaps by multiplying the individual Gaussian viscosity perturbation. However, this is not based on simulating the gas-planet interaction in a common gap of two planets \citep[e.g.][]{masset_reversing_2001, bergez-casalou_simultaneous_2023}. But it is unlikely to affect our conclusions since the exact shape of the gap does not play a role as long as the gap can block pebbles efficiently. However, a more realistic overlapping of gaps can affect the pebble and gas reservoir between planets and may change their accretion \citep{weber_characterizing_2018}. This could be investigated further in future studies. 

Fourth, the disk chemistry is static and molecules can not be converted into other species. In principle, chemical reactions can change the composition of ice and gas. Studies show that depending on the level of ionisation of the disk the chemical evolution becomes significant after a few $10^5$ yr and settles into a steady state after 7 Myr \citep{eistrup_setting_2016, eistrup_molecular_2018}. The chemical evolution of pebbles is found to be slower than their drift \citep{booth_planet-forming_2019, eistrup_chemical_2022}, but can become important if the presence of a pressure gap slows down the inward drift. The latter effect is most likely to influence the results of this study. Especially at low viscosity, we see that large amounts of pebbles are blocked outside the planet's gap and evaporate when the planet crosses an ice line. 
We suggest that future studies investigate the change in the chemical composition of blocked pebbles. 
We note that even if chemical reactions take place,  the overall abundance is unchanged. If the planet migrates into hotter regions these molecules could be accreted by the planet in the gaseous form. It is therefore critical to determine the physical state of the different elements and the associated chemistry in various conditions as well as determining the accretion rates of gas and solids during planet formation.

Fifth, we assume a fixed starting time of $5\times10^4$ years for the planetary embryos. Different starting times affect the growth of planets and change the amount of pebbles that pass the planet before a deep gap is formed. 
Starting at later times leaves more time for pebbles to drift inwards. In this case, the reduction of the initial water enrichment we see in Fig. (\ref{fig:WaterInIceLineAll}) would be smaller.
Also, the late enrichment that occurs after a planet crosses an ice line will be smaller because most pebbles drift away early.
However, our choice of the early starting times can be justified due to the following reasons. First, a planetary embryo starting later than $\sim0.5$ Myr is less likely to form a giant \citep{savvidou_how_2023}. Second, disk substructures indicate an early planet formation \citep{delussu_population_2024}, although not all gaps are caused by planets \citep[e.g.][]{tzouvanou_all_2023}. Clearly, it would be interesting to investigate the sensitivity of the results when considering different starting times and we hope to explore this in future research.  

Sixth, solving eq. (\ref{eq:viscous-evolution}) independently for every gas species assumes there is no angular momentum exchange between the species. This will change gas velocities compared to a case where the gas is fully interacting.
A comparison between different disk models that include multiple molecular species and how to correctly model the diffusion of the species can be found in \citet{desch_formulas_2017}. 
To justify our model, we show the gas velocity of water vapour relative to the gas velocity of the H \& He gas in the appendix \ref{sec:WaterTransport}.
For a short time, the water vapour spreads very fast throughout the disk. 
We can therefore expect that water spreads faster than the viscous time scale because a high composition gradient should lead to fast diffusion through the disk. 
On the other hand, the water gas velocity is lower than the H \& He gas when the water fraction is uniform within the ice line. In that case, we should expect it to be similar to the velocity of the H \& He. 
As shown in \citet{desch_formulas_2017} the methods used here tend to overestimate the diffusion through the disk. However, this effect is only relevant during the early disk evolution ($t \lesssim 1$ Myr) and for disks with a low viscosity.
Within our model, the change in the atmospheric abundances is significant and qualitative trends should be independent of the exact mechanism of gas transport. 
However, when trying to explain the chemical composition and formation scenario of a specific planetary system rather than general trends, a self-consistent treatment of the interactions between different species and their angular momentum exchange would be necessary.

Finally, this study did not include planetesimals although planetesimals are expected to exist. For example, planet gaps with pressure bumps can increase planetesimal formation \citep{eriksson_fate_2021, sandor_planetesimal_2024} and reduce the inward pebble flux \citep{danti_composition_2023}. 
In the case of our model, this will reduce the effect of an outer planet decreasing volatiles in an inner planet's atmosphere by blocking pebbles although gaps blocking pebbles may be more important than planetesimals locking material in place \citep{kalyaan_effect_2023}. 
On the other hand, planetesimals can be swept up by an inward-migrating planet \citep{shibata_origin_2020, shibata_origin_2022} and evaporate at ice lines. Additionally, an inward-migrating planet raises the eccentricities of planetesimals potentially increasing the density of pebbles and dust by ablation and fragmentation \citep{batygin_jupiters_2015, eriksson_fate_2021}.
Furthermore, planetesimals that form in the pressure bump of a planet can be scattered inwards and ablate due to high eccentricities and gas drag \citep{eriksson_fate_2021}. This would decrease the effect of pebble blocking. Accounting for the effect of planetesimals in future research is clearly desirable.

\section{Summary and conclusions}\label{sec:summary}
Planetary systems often harbour multiple planets, hence formation models should include planet formation in the context of other companions rather than focusing on a single planet. We used our planet formation model to investigate the influence of multiple planets on the inner disk and planets by blocking pebbles and preventing enrichment of the disk due to pebble evaporation. Therefore we extended the \chemcomp code, first presented in \citet{schneider_how_2021-1}, to the formation of multiple planets simultaneously.

We showed that gap-opening planets can reduce the water vapour enrichment in the disk. Similar to the study of \citet{kalyaan_effect_2023}, we found that the most important effect originates from the innermost planet outside an ice line. A planet close to the water ice line can grow quickly to block water-rich pebbles and decrease the water vapour content of the inner disk. 
However, the planet is likely to cross the evaporation front, so all the trapped pebbles can evaporate after all resulting in multiple phases of enrichment. The amount of these late enrichment phases depends then on the next planet further out.
The amount of this later enrichment phase depends on the next planet outside the evaporation front.

The effect of blocked pebbles is more important at lower viscosities because pebble evaporation-induced enrichment is stronger, gas transport is slower \citep[e.g.][]{schneider_how_2021, mah_close-ice_2023} and planet migration is reduced. If planets remain longer in outer regions of the disk they can block more pebbles that otherwise would evaporate. Also, we find that in every simulation there is always an initial phase where enough pebbles drift inwards to alter the chemical composition of the inner disk region before planets grow enough to block pebbles. 

The C/O ratio in the disk is mostly affected by the relatively abundant water vapour. 
We also find, that trapping carbon rich pebbles affects the C/O ratio less than blocking water rich pebbles.
In particular, due to the delayed gap opening (caused by the growth of a giant planet), water-rich pebbles can drift inwards and evaporate in all simulations. Therefore, the disks in all our simulations show a phase with a sub-solar C/O ratio inside the water ice line. This implies that young disks ($\lesssim1$ Myr) with super-solar C/O ratios need to harbour initial pressure perturbations that are not caused by growing planets.  

The presence of a gap-opening planet outside the water ice line can reduce the volatile enrichment in the bulk composition of a giant planet's atmosphere inside the water ice line.
In our study, we show that the atmospheric C/H and O/H can be reduced by $\sim30\%$ and $\sim50\%$, respectively, when other planets exist farther out and block volatile-rich pebbles. However, the importance of blocked pebbles depends on the exact scenario, i.e. how planets migrate relative to each other and how they cross the evaporation fronts. Therefore, the  viscosity is a key property in our simulations. 
Yet, the presence of an outer planet generally does not imply that the inner planet has sub-solar C/H or O/H.
Overall this effect is less important than expected because large amounts of pebbles can drift inwards relatively early before planets can carve a pebble-blocking gap. Additionally, when planets migrate over evaporation fronts the trapped material evaporates and enriches the gas again, which can be accreted by an inner planet.

In all our simulations the bulk atmospheric abundance of nitrogen was found to be super-solar. 
We assume that $10\%$ of nitrogen is contained in NH$_3$ and the rest in N$_2$. Because the N$_2$ evaporation front is far out we need to distinguish between two cases. First, the viscous evolution is fast enough to transport N$_2$ gas into the inner disk region. Second, the evolution is too slow. In the first case blocking NH$_3$ rich pebbles can reduce the nitrogen abundance in the atmosphere of close-in planets by $\sim25\%$. In contrast, in the second case this reduction is only $\sim15\%$ because N$_2$ enriched gas can be accreted instead. However, if the outermost planet is initially located beyond the N$_2$ evaporation front, this enrichment can be reduced as well. 
In both cases, our model implies that exoplanetary atmospheres should have super-stellar nitrogen abundances if pebble drift and evaporation are significant.
We note that studies that focused on nitrogen enrichment in Jupiter's atmosphere suggested that this enrichment could also be a result of core formation near or beyond the N$_2$ ice line \citep{bosman_jupiter_2019, oberg_jupiters_2019}.  However, in our case nitrogen  enrichment is not necessarily a result of formation near the 
N$_2$ ice line making it a more common outcome of planet formation. 
In any case, we can conclude that nitrogen is an important element in constraining the planetary formation history and hope to address this topic in future research.

Our study suggests that the number of planets can affect the formation scenario of individual planets and plays a crucial role. However, there are improvements to enhance the model that should be made. Additionally, a broader parameter space will give more insight into the complex interplay between multiple growing planets. Further investigations of these topics are required in order to link the atmospheric composition of exoplanets with their formation location.

\begin{acknowledgements}
We thank Sho Shibata for the helpful discussion on the effects of planetesimals and Christoph Mordasini for a helpful discussion. ME and RH acknowledge support from SNSF grant \texttt{\detokenize{ 	200020_215634}}. 
We also thank the referee for the comments that improved the quality of our manuscript.
\end{acknowledgements}

\bibliographystyle{aa.bst}
\bibliography{paper.bib}

\begin{appendix}
\section{Pebble size limits}\label{sec:PebbleSizeLimits}

\begin{figure*}
    \centering
    \includegraphics[width=0.97\linewidth]{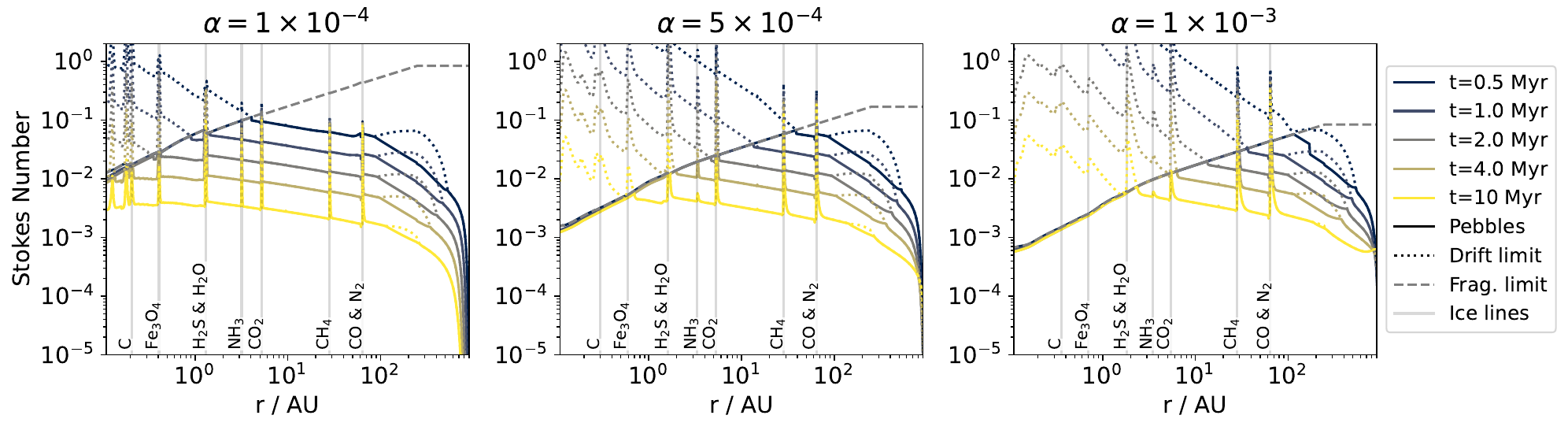}
    \caption{Stokes Number of pebbles (solid lines), the drift limit (dotted lines) and fragmentation limit (grey dashed line) at different times (colours) for the unperturbed disk. The different viscosities are shown from left to right (setups: V1noP, V5noP \& V10noP). Vertical lines show the position of various ice lines. The fragmentation limit is only shown for $t=0.5$ Myr because it is almost constant over time.
    \label{fig:StokesNumber}}
\end{figure*}

We follow the two-population model of \citet{birnstiel_simple_2012} to determine the pebble sizes. The Stokes number St in the Epstein drag regime is given by
\begin{equation}
    \text{St} = \Omega_\text{K}\tau_f = \frac{\pi a\rho_\bullet}{2\Sigma_\text{gas}}
\end{equation}
with $\tau_f$, $a$ and $\rho_\bullet$ being the friction time, particle size and density respectively. The pebble size limits can be reformulated in terms of a maximum Stokes number. The drift limit St$_\text{dritf}$, the fragmentation limit St$_\text{frag}$ and the relative drift-induced fragmentation limit St$_\text{df}$, are given by the equations:
\begin{align}
    \text{St}_\text{drift}&=f_\text{d}\frac{v_\text{K}^2}{\gamma c_\text{s}^2}\frac{\Sigma_\text{dust}}{\Sigma_\text{gas}}\\
    \text{St}_\text{frag}&=f_\text{f} \frac{v_\text{f}^2}{3\alpha c_\text{s}^2}\\
    \text{St}_\text{df}&=\frac{v_\text{f}v_\text{K}^2}{\gamma c_\text{s}^2 (1-N)}
\end{align}
with $v_\text{K}=r\Omega_\text{K}$ being the Kepler velocity, $\gamma=\lvert\frac{\text{d}\ln P}{\text{d}\ln r}\rvert$ the logarithmic pressure gradient, $\Sigma_\text{solid}$ the dust surface density, $f_\text{d}=0.55$ and $f_\text{f}=0.37$ fit factors and $N=0.5$ a factor that expresses the size differences of particles that collide due to different drift speeds. In Fig. (\ref{fig:StokesNumber}) we show the Stokes number of pebbles and the limiting Stokes numbers at different times in the disk. We exclude the drift-induced fragmentation limit from the plot because it's always higher than the other two limiting processes.

The fragmentation limit is almost constant in time because the only changing quantity is $c_\text{s}$. We assume a time-independent temperature profile but the mean molecular weight of the gas can change $c_\text{s}$. However, the mean molecular weight is dominated by hydrogen and helium and therefore almost constant over time.

For the low viscosity, pebbles just outside the water ice line are limited by fragmentation for $t\lesssim1$ Myr. So early on, pebbles can grow exponentially to large sizes and enrich the inner disk with high amounts of water. After $t\sim1$\,Myr pebbles outside the water ice-line become fully drift-limited and smaller over time. For the high viscosity case, fragmentation dominates from the inner disk edge to $r\sim5$ AU for as long as $t\lesssim4$\,Myr, resulting in smaller pebbles because the fragmentation limit is inversely proportional to $\alpha$. The drift limit is only important further out.

\section{Water transport}\label{sec:WaterTransport}
\begin{figure*}[htb!]
    \centering
    \includegraphics[width=0.99\linewidth]{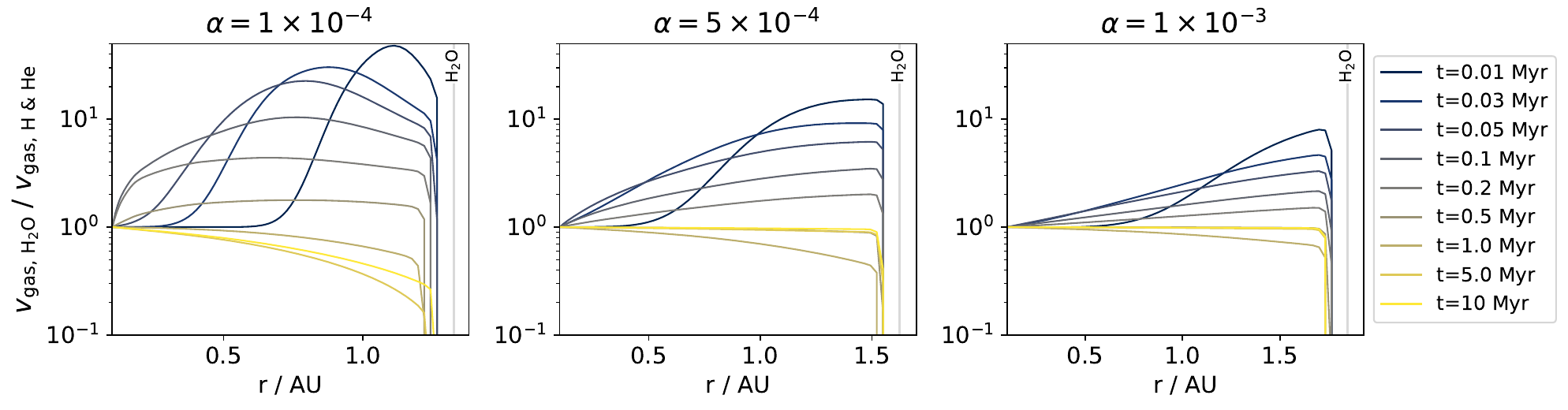}
    \caption{Velocity of the water vapour relative to the velocity of H \& He at different times (colours) for the unperturbed disk. The different viscosities are shown from left to right (setups: V1noP, V5noP \& V10noP). The vertical line shows the position of the water ice lines. Note a positive velocity ratio indicates an inward flow of gas. Note positive values indicate inward movement.
    \label{fig:WaterVelocity}}
\end{figure*}
We show the water gas velocity relative to the  H \& He gas velocity in Fig. (\ref{fig:WaterVelocity}). 
In our model, each molecular species can have different velocities because we solve eq. (\ref{eq:viscous-evolution}) for each species individually.
When pebbles evaporate they create a density enhancement in the gas leading to a diffusion of water faster than the viscous time scale of the disk.
Hence, the water is very fast initially and quickly increases the water fraction towards the inner edge of the gas disk. The water velocity becomes slower than the velocity of the H \& He gas when the water fraction becomes uniform across the inner region of the disk (around $t\sim1$\,Myr for the low viscosity).
Close to the water ice line (within $\lesssim0.1$ AU from its location) the velocity decreases and gets negative to ensure the conservation of angular momentum within the water vapour.
Water vapour is transported outward, where it re-condensates after crossing the ice line.
As discussed in the caveats the diffusion of water vapour is overestimated early in the disk. For a comparison between different viscous evolution equations that treat the transport of individual molecular species correctly see \citet{desch_formulas_2017}.

\begin{figure}[H]
    \centering
    \includegraphics[width=0.99\linewidth]{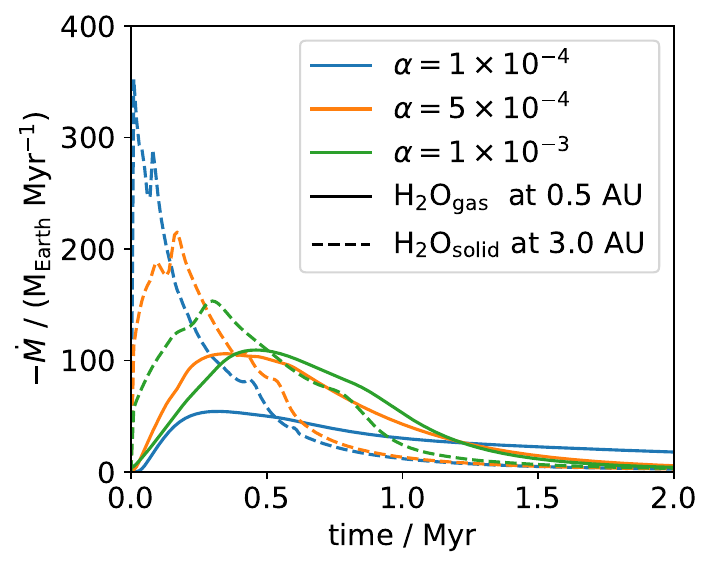}
    \caption{Disk accretion rate $\dot{M}$ of water gas at 0.5 AU (solid lines) and water solids at 3.0 AU (dashed lines) over time of the unperturbed disk. Different colours represent the different viscosities (setups: V1noP, V5noP \& V10noP). The water ice line is between 1 and 2 AU depending on the viscosity. Note positive values indicate inward movement.
    \label{fig:WaterDiskAccretionRate}}
\end{figure}
Fig. (\ref{fig:WaterDiskAccretionRate}) shows the disk accretion rate $\dot{M}$ of water vapour at 0.5 AU (inside the water ice line) and the disk accretion rate of solids at 3 AU (outside the water ice line). Initially, pebble drift carries more water towards the ice line than water vapour is transported away by viscosity. This increases the water fraction of the gas inside the water ice line. With higher viscosity, gas transport is more efficient, quickly reducing this excess water. At low viscosity, the gas transport is slow and the solid accretion rate is high because pebbles are larger. This leads to a high water fraction in the gas disk that persists for a long time. In contrast, at high viscosity, the solid accretion rate increases slower and reaches only $\sim50\%$ of the low viscosity maximum value. However, gas transport is more efficient, resulting in a lower and shorter-lived increase in the water fraction of the gas disk (see Sec.~\ref{sec:results}  \&  Fig.~(\ref{fig:WaterHeatmapAll})).

\section{Disk nitrogen abundance}\label{sec:DiskNitrogenAbundance}
\begin{figure*}[htb!]
    \centering
    \includegraphics[width=0.99\linewidth]{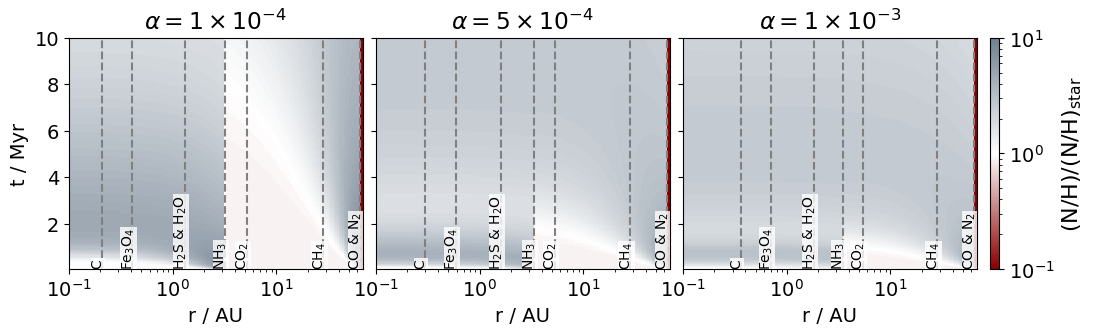}
    \caption{Gas N/H number ratio normalised to the solar value as a function of time and radius. The three viscosities are shown from left to right (setups: V1noP, V5noP \& V10noP).
    \label{fig:DiskNitrogenAbundance}}
\end{figure*}
We show the N/H ratio normalised by the stellar value of the unperturbed disk in Fig. (\ref{fig:DiskNitrogenAbundance}). At low viscosity, the nitrogen contained in NH$_3$ rich pebbles increases the abundance to super stellar values in the inner disk regions. The ratio stays high for the entire disk lifetime. Gas enriched by N$_2$ containing pebbles is too slow to reach radii of $r\lesssim3$ AU within the 10 Myr, making the region outside the methane ice line less enriched for most of the disk's lifetime. At higher viscosity, N$_2$ enriched gas can reach the inner region within the disk lifetime, similar to methane. Consequently, the whole disk has a super stellar nitrogen abundance for most of the disk's lifetime.

\section{Final planet composition and mass.}\label{sec:FinalPlanetCompositions}
\begin{table*}[hb]
    \caption{Final planet composition and mass.}
    \label{tab:Simulation_Results}

\begin{tabular}{|l|c|cccc|cccc|cc|}
\hline
Setup & Planet & C/H & O/H & N/H & C/O & \{C/H\}$_\text{s}$ & \{O/H\}$_\text{s}$ & \{N/H\}$_\text{s}$ & \{C/O\}$_\text{s}$ & $M$ [$M_\text{J}$] & $M_\text{env}/M$ [\%] \\
\hline\hline\multicolumn{12}{|c|}{$\alpha=1\times10^{-4}$}\\\hline
V1P1 & 1 & 2.85 & 9.38 & 2.48 & 0.30 & 1.00 & 1.00 & 1.00 & 1.00 & 2.96 & 99.43 \\
\hline V1P2 & 2 & 2.06 & 0.75 & 1.21 & 2.75 & 1.00 & 1.00 & 1.00 & 1.00 & 1.25 & 95.78 \\
\hline V1P3 & 3 & 7.90 & 10.11 & 3.01 & 0.78 & 1.00 & 1.00 & 1.00 & 1.00 & 0.20 & 50.79 \\
\hline\multirow{2}{*}{V1P12} & 1 & 2.25 & 4.25 & 1.82 & 0.53 & 0.79 & 0.45 & 0.73 & 1.74 & 3.22 & 99.48 \\
                             & 2 & 2.11 & 0.70 & 1.26 & 3.02 & 1.02 & 0.93 & 1.04 & 1.10 & 1.48 & 96.42 \\
\hline\multirow{2}{*}{V1P13} & 1 & 2.51 & 7.31 & 2.21 & 0.34 & 0.88 & 0.78 & 0.89 & 1.13 & 2.95 & 99.43 \\
                             & 3 & 6.53 & 8.15 & 2.93 & 0.80 & 0.83 & 0.81 & 0.97 & 1.02 & 0.22 & 56.11 \\
\hline\multirow{2}{*}{V1P23} & 2 & 1.84 & 0.60 & 1.27 & 3.05 & 0.89 & 0.80 & 1.05 & 1.11 & 2.01 & 97.39 \\
                             & 3 & 2.66 & 2.80 & 2.36 & 0.95 & 0.34 & 0.28 & 0.79 & 1.22 & 0.50 & 80.48 \\
\hline\multirow{3}{*}{V1P123} & 1 & 2.12 & 3.99 & 1.82 & 0.53 & 0.74 & 0.43 & 0.73 & 1.75 & 3.38 & 99.50 \\
                              & 2 & 1.87 & 0.59 & 1.32 & 3.19 & 0.91 & 0.78 & 1.09 & 1.16 & 2.25 & 97.67 \\
                              & 3 & 2.66 & 2.79 & 2.37 & 0.95 & 0.34 & 0.28 & 0.79 & 1.22 & 0.50 & 80.56 \\

\hline\hline\multicolumn{12}{|c|}{$\alpha=5\times10^{-4}$}\\\hline
V5P1 & 1 & 4.13 & 4.68 & 2.52 & 0.88 & 1.00 & 1.00 & 1.00 & 1.00 & 5.46 & 99.69 \\
\hline V5P2 & 2 & 2.66 & 2.36 & 2.39 & 1.13 & 1.00 & 1.00 & 1.00 & 1.00 & 6.88 & 99.34 \\
\hline V5P3 & 3 & 2.49 & 0.94 & 2.42 & 2.65 & 1.00 & 1.00 & 1.00 & 1.00 & 6.51 & 99.05 \\
\hline\multirow{2}{*}{V5P12} & 1 & 2.93 & 3.19 & 2.24 & 0.92 & 0.71 & 0.68 & 0.89 & 1.04 & 4.48 & 99.62 \\
                             & 2 & 2.60 & 2.10 & 2.29 & 1.23 & 0.98 & 0.89 & 0.96 & 1.09 & 6.60 & 99.32 \\
\hline\multirow{2}{*}{V5P13} & 1 & 3.54 & 3.66 & 2.33 & 0.97 & 0.86 & 0.78 & 0.92 & 1.10 & 4.89 & 99.65 \\
                             & 3 & 2.45 & 0.88 & 2.32 & 2.77 & 0.98 & 0.94 & 0.96 & 1.05 & 6.77 & 99.09 \\
\hline\multirow{2}{*}{V5P23} & 2 & 2.35 & 0.97 & 1.98 & 2.42 & 0.89 & 0.41 & 0.83 & 2.15 & 5.71 & 99.22 \\
                             & 3 & 2.42 & 0.82 & 2.24 & 2.93 & 0.97 & 0.88 & 0.93 & 1.11 & 6.76 & 99.08 \\
\hline\multirow{3}{*}{V5P123} & 1 & 2.76 & 2.37 & 2.18 & 1.16 & 0.67 & 0.51 & 0.87 & 1.32 & 4.35 & 99.61 \\
                              & 2 & 2.21 & 0.83 & 1.85 & 2.67 & 0.83 & 0.35 & 0.77 & 2.36 & 5.67 & 99.22 \\
                              & 3 & 2.30 & 0.70 & 2.24 & 3.30 & 0.93 & 0.74 & 0.92 & 1.25 & 6.73 & 99.08 \\

\hline\hline\multicolumn{12}{|c|}{$\alpha=1\times10^{-3}$}\\\hline
V10P1 & 1 & 3.35 & 3.44 & 2.40 & 0.97 & 1.00 & 1.00 & 1.00 & 1.00 & 5.23 & 99.60 \\
\hline V10P2 & 2 & 2.49 & 2.61 & 2.27 & 0.95 & 1.00 & 1.00 & 1.00 & 1.00 & 5.57 & 99.28 \\
\hline V10P3 & 3 & 2.30 & 1.75 & 2.22 & 1.31 & 1.00 & 1.00 & 1.00 & 1.00 & 5.98 & 99.12 \\
\hline\multirow{2}{*}{V10P12} & 1 & 3.28 & 3.57 & 2.25 & 0.92 & 0.98 & 1.04 & 0.94 & 0.94 & 3.46 & 99.40 \\
                              & 2 & 2.39 & 2.45 & 2.24 & 0.98 & 0.96 & 0.94 & 0.99 & 1.02 & 5.43 & 99.26 \\
\hline\multirow{2}{*}{V10P13} & 1 & 3.35 & 3.65 & 2.29 & 0.92 & 1.00 & 1.06 & 0.95 & 0.94 & 3.80 & 99.45 \\
                              & 3 & 2.25 & 1.66 & 2.17 & 1.36 & 0.98 & 0.94 & 0.98 & 1.04 & 5.68 & 99.07 \\
\hline\multirow{2}{*}{V10P23} & 2 & 2.46 & 2.24 & 2.10 & 1.10 & 0.99 & 0.86 & 0.92 & 1.15 & 4.63 & 99.14 \\
                              & 3 & 2.22 & 1.60 & 2.20 & 1.38 & 0.97 & 0.91 & 0.99 & 1.06 & 5.91 & 99.11 \\
\hline\multirow{3}{*}{V10P123} & 1 & 2.94 & 2.97 & 2.26 & 0.99 & 0.88 & 0.86 & 0.94 & 1.01 & 3.75 & 99.45 \\
                               & 2 & 2.23 & 2.08 & 2.04 & 1.07 & 0.89 & 0.80 & 0.90 & 1.12 & 4.41 & 99.10 \\
                               & 3 & 2.19 & 1.51 & 2.19 & 1.45 & 0.95 & 0.86 & 0.98 & 1.10 & 5.86 & 99.11 \\
\hline
\end{tabular}

    \tablefoot{The chemical abundance ratios in the atmosphere of the planets are normalised by the solar value ($\cdot/\cdot$). Additionally, we show the chemical abundance ratios in the atmosphere of the planets relative to the single planet case ($\{\cdot/\cdot\}_\text{s}$). Finally, we show the total mass of the planet $M$ and the fraction of the planet's envelope mass to the planet's total mass $M_\text{env}/M$. The envelope mass is defined as the mass accreted during the gas accretion phase plus 10\% of the accreted solid material as described in the Sec. (\ref{sec:method}).}

\end{table*}

\clearpage
\FloatBarrier
\section{Condensation temperatures and volume mixing ratios}\label{sec:CondensationTAndVolMixRatios}
\begin{table}[!htb]
\caption{Condensation temperatures and volume mixing ratios (by number).}
\label{tab:mixing_ratios}
    \centering
        \begin{tabular}{|c|c|>{\centering\arraybackslash}p{5.0cm}|}
        \hline
        Species & $T\low{cond}$       & volume mixing ratio\\\hline
        CO		    & 20		      & 0.45 $\times$ C/H \\ 
        N$_2$		& 20		      & 0.45 $\times$ N/H  \\ 
        CH$_4$		& 30		      & 0.25 $\times$ C/H \\ 
        CO$_2$		& 70		      & 0.1 $\times$ C/H  \\ 
        NH$_3$		& 90		      & 0.1 $\times$ N/H  \\ 
        H$_2$S		& 150		      & 0.1 $\times$ S/H  \\ 
        H$_2$O		& 150		      & O/H - (3 $\times$ MgSiO$_3$/H + 4 $\times$ Mg$_2$SiO$_4$/H + CO/H + 2 $\times$ CO$_2$/H + 3 $\times$ Fe$_2$O$_3$/H + 4 $\times$ Fe$_3$O$_4$/H + VO/H + TiO/H + 3$\times$Al$_2$O$_3$ + 8$\times$NaAlSi$_3$O$_8$ + 8$\times$KAlSi$_3$O$_8$)\\ 
        Fe$_3$O$_4$	& 371		      & (1/6) $\times$ (Fe/H - 0.9 $\times$ S/H) \\ 
        C (grains)   & 631	  & 0.2 $\times$ C/H  \\ 
        FeS		    & 704		      & 0.9 $\times$ S/H  \\ 
        NaAlSi$_3$O$_8$	    & 958	  & Na/H 		       \\ 
        KAlSi$_3$O$_8$		& 1006	  & K/H 		       \\ 
        Mg$_2$SiO$_4$		& 1354	  & Mg/H - (Si/H - 3$\times$K/H - 3$\times$Na/H) \\ 
        Fe$_2$O$_3$	& 1357		      & 0.25 $\times$ (Fe/H - 0.9 $\times$ S/H) \\ 
        VO		    & 1423		      & V/H 		      \\ 
        MgSiO$_3$	& 1500		      & Mg/H - 2$\times$(Mg/H - (Si/H - 3$\times$K/H - 3$\times$Na/H)) 		                           \\ 
        Al$_2$O$_3$	& 1653		      & 0.5$\times$(Al/H - (K/H + Na/H)) 		\\ 
        TiO		   & 2000		      & Ti/H \\\hline
        \end{tabular}
    \tablefoot{We use the 20\% carbon grain model of \citet{schneider_how_2021-1} for the mixing ratios. Condensation temperatures of molecules are taken from \citet{lodders_solar_2003}. We assume Fe$_2$O$_3$ has the same condensation temperature as Fe.}
\end{table}

\section{Initial disk composition}\label{sec:InitialDiskComposition}
\begin{table}[htb!]
\caption{Initial disk composition based on solar elemental abundances \citep{asplund_chemical_2009}}
\label{tab:solar_elemental_abundances}
    \centering
    \begin{tabular}{|c|c|}
        \hline
        Species & Abundance\\\hline
        He/H & 0.085\\
        O/H & 4.90 $\times$ 10$^{-4}$\\
        C/H & 2.69 $\times$ 10$^{-4}$\\
        N/H & 6.76 $\times$ 10$^{-5}$\\
        Mg/H & 3.98 $\times$ 10$^{-5}$\\
        Si/H & 3.24 $\times$ 10$^{-5}$\\
        Fe/H & 3.16 $\times$ 10$^{-5}$\\
        S/H & 1.32 $\times$ 10$^{-5}$\\
        Al/H & 2.82 $\times$ 10$^{-6}$\\
        Na/H & 1.74 $\times$ 10$^{-6}$\\
        K/H & 1.07 $\times$ 10$^{-7}$\\
        Ti/H & 8.91 $\times$ 10$^{-8}$\\
        V/H & 8.59 $\times$ 10$^{-9}$\\\hline
        \end{tabular}
\end{table}

\end{appendix}
\end{document}